\documentclass[fleqn,usenatbib]{mnras}

\usepackage[T1]{fontenc}
\usepackage{ae,aecompl}

\usepackage{graphicx}             
\usepackage{amsmath}           
\usepackage{amssymb}           
\usepackage{times}                  
\usepackage{enumitem}           
\usepackage{xcolor}                 
\usepackage{longtable}         
\usepackage{lscape}
\usepackage{caption}
\usepackage{calc}

\pdfoutput=1

\newcommand{\Mjup}{\mbox{$M_{\rm Jup}$}}

\newcommand{\degs}{\mbox{$^{\circ}$}}

\newcommand{\eg}{e.g.\,\,}
\newcommand{\ie}{i.e.\,\,}

\newcommand{\Teff}{\mbox{$T_{\rm eff}$}\,}
\newcommand{\Mpc}{\mbox{Mpc}}

\newcommand{\mum}{\mbox{$ \rm \mu$m}}

\newcommand{\refequ}[1]{equation (\ref{#1})}

\newcommand{\arcseconds}{\,\mbox{arcsec}\,}

\newcommand{\RZ}{\mbox{$(r-z)$}\,}

\newcommand{\VJ}{\mbox{$(V-J)$}\,}
\newcommand{\JH}{\mbox{$(J-H)$}\,}
\newcommand{\JK}{\mbox{$(J-K_S)$}\,}

\newcommand{\JWa}{\mbox{$(J-W1)$}\,}
\newcommand{\JWb}{\mbox{$(J-W2)$}\,}
\newcommand{\HWa}{\mbox{$(H-W1)$}\,}
\newcommand{\HWb}{\mbox{$(H-W2)$}\,}
\newcommand{\KWb}{\mbox{$(K_S-W2)$}\,}

\newcommand{\WaWc}{\mbox{$(W1-W3)$}\,}
\newcommand{\WbWc}{\mbox{$(W2-W3)$}\,}

\newcommand{\JKcut}{\mbox{$Y/K$}\,}

\defcitealias{Cook2016}{paper 1}

\title[Low-res NIR follow-up of M+UCDs]{Low-resolution near-infrared spectroscopic signatures of unresolved ultracool companions to M dwarfs}
\author[N. J. Cook et~al.]{N. J. ~Cook,\thanks{E-mail: \href{mailto:neil.james.cook@gmail.com}{neil.james.cook@gmail.com}}$^{1, 2}$
D. J. ~Pinfield,$^{2}$ F. ~Marocco,$^{2}$ B. ~Burningham,$^{2, 3}$ H. R. A. ~Jones,$^{2}$ 
\newauthor
J. ~Frith,$^{2}$ J. ~Zhong,$^{4}$ A. L. ~Luo,$^{5}$ Z. X. ~Qi,$^{5}$ N. B. Cowan,$^{6, 7, 8, 9}$ M. Gromadzki,$^{10}$ 
\newauthor
R. G. ~Kurtev,$^{11, 12}$ Y. X. ~Guo,$^{5}$ Y. F. ~Wang,$^{5}$ Y. H. ~Song,$^{5}$ Z. P. ~Yi,$^{13}$ and R. L. ~Smart$^{14}$
\\\\
$^{1}$SC-Physics \& Astronomy, Petrie Science \& Engineering, 4700 Keele Street, Toronto, ON, M3J1P3 \\
$^{2}$Centre for Astrophysics Research, School of Physics, Astronomy, and Mathematics, University of Hertfordshire, College Lane, Hatfield AL10 9AB, UK \\
$^{3}$NASA Ames Research Center, Mail Stop 245-3, Moffett Field, CA 94035, USA\\
$^{4}$Key Laboratory for Research in Galaxies and Cosmology, SHAO, Chinese Academy of Sciences, 80 Nandan Road, Shanghai 200030, China\\ 
$^{5}$Key Laboratory of Optical Astronomy, NAO, Chinese Academy of Sciences, Datun Road 20A, Beijing 100012, China\\
$^{6}$Department of Earth \& Planetary Sciences, McGill University, 3450 rue University, Montr$\acute{e}$al, QC, H3A 0E8, Canada\\
$^{7}$Department of Physics, McGill University, 3600 rue University, Montr$\acute{e}$al, QC, H3A 2T8, Canada\\
$^{8}$McGill Space Institute, 3550 University Street, Montreal, QC, H3A 2A7, Canada\\
$^{9}$Institut de recherche sur les exoplan$\grave{e}$tes, D$\acute{e}$partement de physique, Universit$\acute{e}$ de Montr$\acute{e}$al, Montr$\acute{e}$al, QC, H3C 3J7, Canada\\
$^{10}$Warsaw University Astronomical Observatory, Al. Ujazdowskie 4, 00-478 Warszawa, Poland\\
$^{11}$Millennium Institute of Astrophysics, Av. Vicua Mackenna 4860, 782-0436, Macul, Santiago, Chile\\
$^{12}$Istitiuto de F{\'i}sica y Astronom{\'i}a, Universidad de Valpara{\'i}so, ave. Gran Breta\~{n}a, 1111, Casilla 5030, Valpara{\'i}so, Chile\\
$^{13}$Shandong University at Weihai,  Weihai, 264209, China\\
$^{14}$Istituto Nazionale di Astrofisica - Osservatorio Astrofisico di Torino, Via Osservatorio 20, I-10023, Torino\\
}

\date{Accepted 2017 February 1. Received 2017 January 18; in original form 2016 August 31}

\pubyear{2017}

\begin{document}
\label{firstpage}
\pagerange{\pageref{firstpage}--\pageref{lastpage}}

\include{aas_macros}

\maketitle

\begin{abstract}
We develop a method to identify the spectroscopic signature of unresolved L-dwarf ultracool companions, which compares the spectra of candidates and their associated control stars using spectral ratio differences and residual spectra. We present SpeX prism-mode spectra (0.7-2.5 $\mum$) for a pilot sample of 111 mid M dwarfs, including 28 that were previously identified as candidates for unresolved ultracool companionship (a sub-sample from Cook et al. 2016; paper 1) and 83 single M dwarfs that were optically colour-similar to these candidates (which we use as `control stars'). We identify four candidates with evidence for near-infrared excess. One of these (WISE J100202.50+074136.3) shows strong evidence for an unresolved L dwarf companion in both its spectral ratio difference and its residual spectra, two most likely have a different source for the near-infrared excess, and the other may be due to spectral noise. We also establish expectations for a null result (i.e. by searching for companionship signatures around the M dwarf control stars), as well as determining the expected outcome for ubiquitous companionship (as a means of comparison with our actual results), using artificially generated unresolved M+L dwarf spectra. The results of these analyses are compared to those for the candidate sample, and reasonable consistency is found. With a full follow-up programme of our candidates sample from Cook et al., we might expect to confirm up to 40 such companions in the future, adding extensively to the known desert population of M3--M5 dwarfs.
\end{abstract}


\section{Introduction}
    \label{section:intro}

\begin{figure*}
\begin{center}
\begin{minipage}{.54\textwidth}
	\begin{center}
	\includegraphics[width=\textwidth]{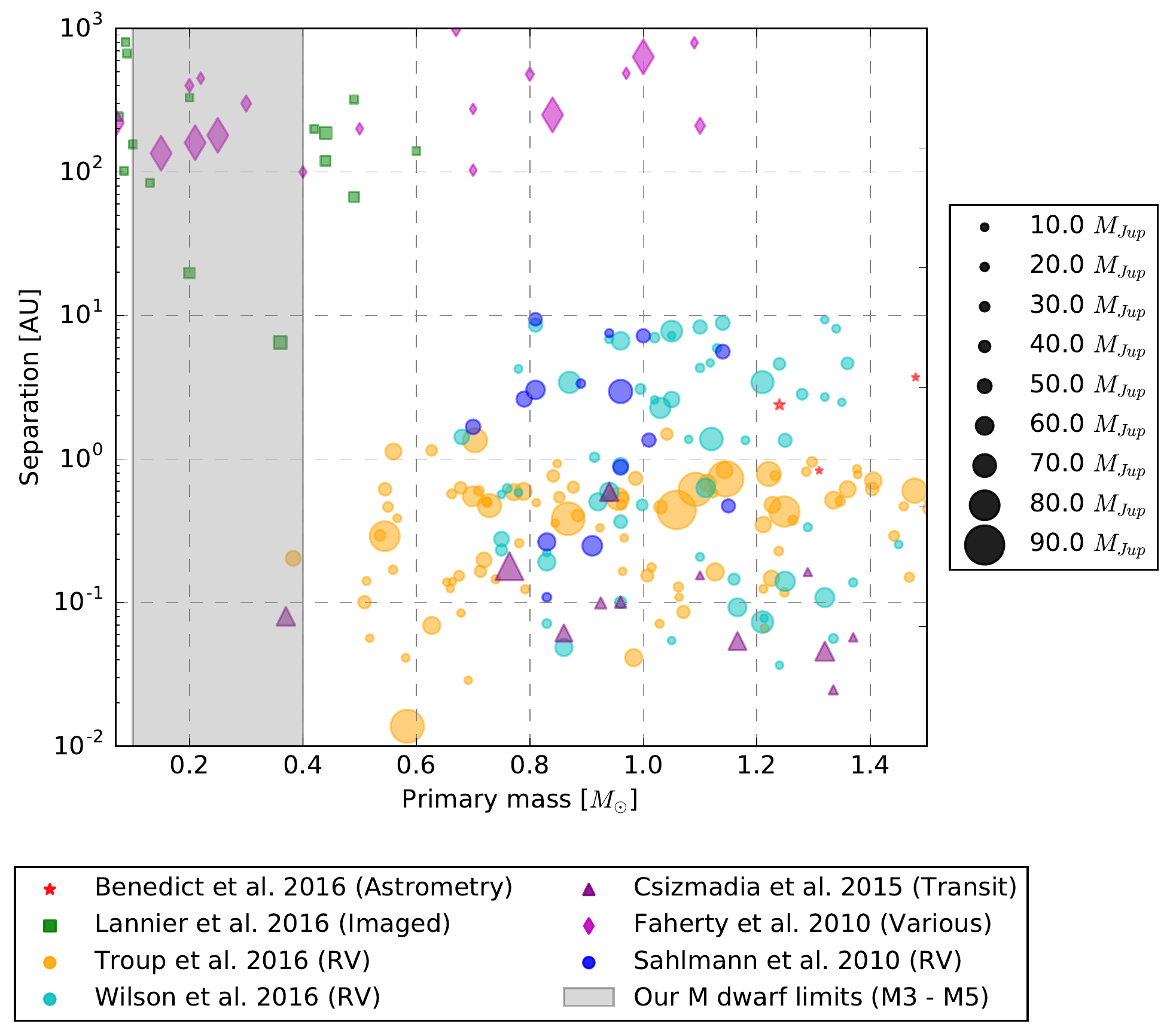}
	\end{center}
\end{minipage}
\begin{minipage}{.435\textwidth}
	\begin{center}
	\includegraphics[width=\textwidth]{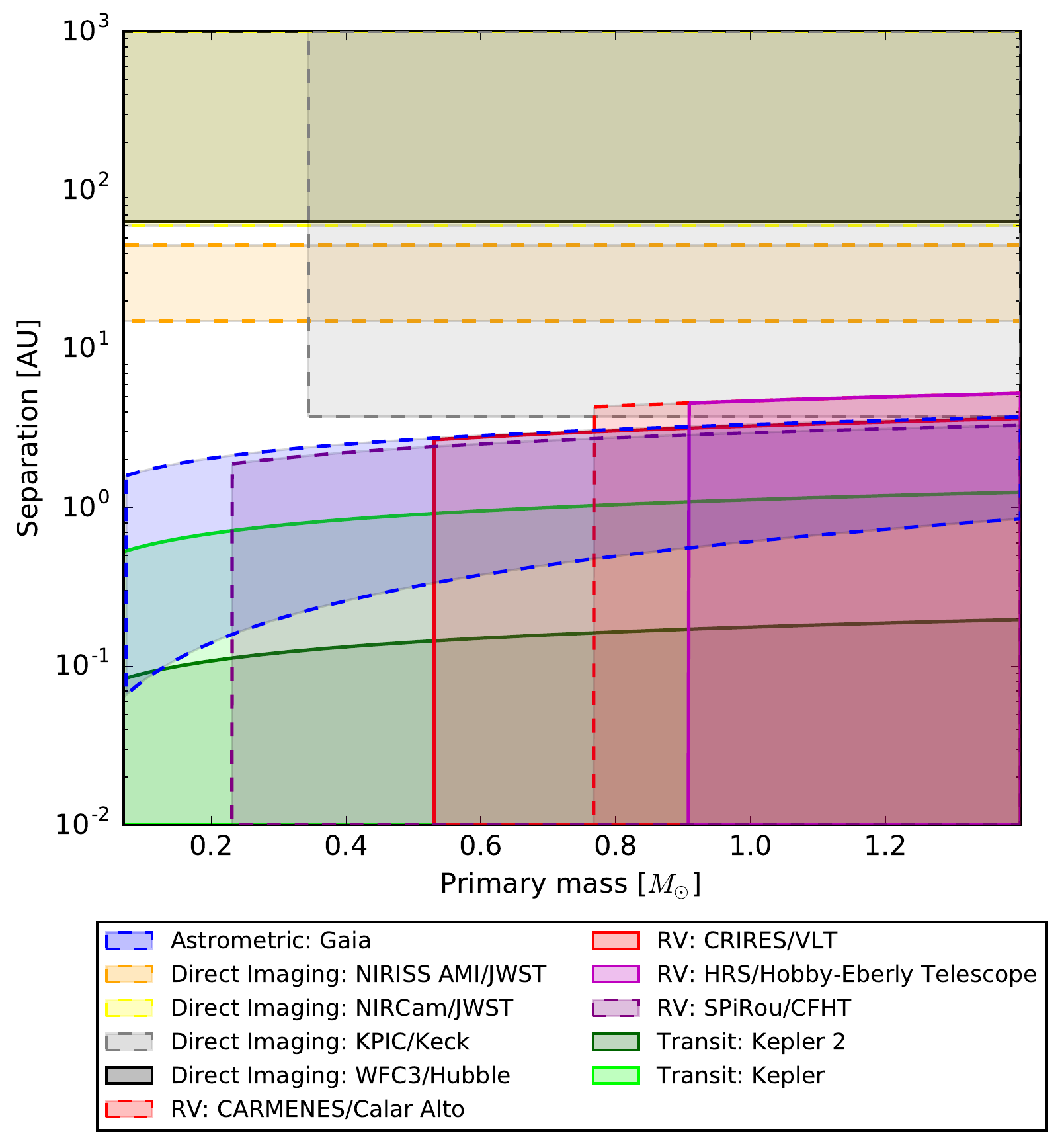}
	\end{center}
\end{minipage}
\end{center}
\caption{Left: a representative sample of known UCD companions to low-mass stars, in the separation range $<$1000 au. Different discovery methods are indicated by different symbols, with symbol size scaling with companion mass (see plot keys). Right: estimated sensitivity regions for different techniques/facilities. We assume the following sensitivity limits: Spatial resolution for {\itshape Hubble}/WFC3, {\itshape JWST}/NIRISS AMI, {\itshape JWST}/NIRCam,Keck/KPIC of 0.4, 0.1--0.3, 0.03, 0.4 \arcseconds; radial velocity limits for HRS, CRIRES, SPiRou, CARMENES of 3, 5, 4, 1 ms$^{-1}$ (SNR$\sim$100), and baselines for these facilities of 10, 6, 5, 3 yr, respectively; astrometric accuracy for {\itshape Gaia} of 150 $\mu as$ over a 6 yr mission; light curve accuracy (for transit detection) of 250 and 300 ppm for Kepler and K2 over 3.5 yr and 80 d baselines, respectively. \label{figure:mass_vs_separation}}
\end{figure*}

Brown dwarf companions to main-sequence stars are of interest for our understanding of star and brown dwarf formation, as well as for the measurement of brown dwarf properties. The `brown dwarf desert' was first identified by radial velocity surveys (\eg \citealt{Marcy2000}) that showed about 5 per cent of solar-type stars have planets ($<13 \Mjup$) within $\sim$5 au, but fewer than 1 per cent of these stars have more massive substellar companions (13--80 \Mjup) in this separation range. The `desert' actually extends up to very low-mass stellar companions ($\sim$100 \Mjup), but disappears at higher companion masses for which the frequency is $\sim$10 per cent \citep{Duquennoy1991,Halbwachs2003}. Further study has shown that the desert covers separation ranges out to several hundred au (\eg \citealt{Gizis2001}; \citealt{McCarthy2004}; \citealt{Cheetham2015}), and also encompasses M dwarfs as well as solar type stars \citep{Dieterich2012}.

The existence of the desert provides an important test for formation models, with a range of factors potentially contributing to its existence. \citet{Jumper2013} suggest turbulent fragmentation alone may give rise to the desert. Alternatively many brown dwarfs may form in massive circumstellar discs, which only undergo primary fragmentation in their cooler outer parts (\citealt{Whitworth2006}, \citealt{Stamatellos2009}; \citealt{Li2015}) leading to a desert at closer separation. It has also been suggested close-in brown dwarfs in a proto-planetary disc will undergo inward migration and destruction via a merger with the star \citep{Armitage2002}.

Detailed study of the desert is hampered by the paucity of brown dwarfs, though a desert population has begun to emerge from studies employing radial velocity and astrometry (\eg \citealt{Wilson2016}), high resolution imaging (\eg \citealt{Kraus2011}; \citealt{Dieterich2012}; \citealt{Hinkley2015}; \citealt{Mawet2015}), microlensing (\eg \citealt{Han2016}) and transit detection \citep{Csizmadia2015}. Indeed, at close separation the large amount of radial velocity data from exoplanet searches is yielding a more detailed picture (\eg \citealt{DeLee2013}; \citealt{Ma2014}), however at wider separations there are still statistically low numbers of companions.

Desert companions are ultracool dwarfs (UCDs; $\gtrsim$M8--M9 type and later, $\lesssim$2500K; \citealt{Chabrier2007}), with their spectral type dependant on mass and age (\eg see Fig. 8 in \citealt{BurrowsHubbardLunineEtAl2001}). Mid-L dwarfs and cooler objects are all substellar. Early L dwarfs may be low-mass stars older than $\sim$2 Gyr, high-mass brown dwarfs with an age $\sim$1--2 Gyr, or younger lower mass brown dwarfs. Late M dwarfs may be low-mass stars (with ages of $\sim$0.2--1 Gyr or greater) or younger brown dwarfs.

In this paper we continue our efforts to identify unresolved UCD companions to M dwarfs from the {\itshape Sloan Digital Sky Survey} \citep[SDSS;][]{York2000}, {\itshape Two Micron All-Sky Survey} \citep[2MASS;][]{Skrutskie2006} and {\itshape Wide-Field Infrared Survey Explorer} \citep[WISE;][]{Wright2010} compilation of \citet[][henceforth \citetalias{Cook2016}]{Cook2016}. Our sample is sensitive to projected separations $\lesssim$450 au (distances of $\sim$150 pc at $\sim$3 \arcseconds resolution), thus spanning the brown dwarf desert. Fig. 1 shows the observational separation versus primary mass plane for ultracool desert companions. In the left-hand panel, known companions are shown, with different discovery methods and published sources indicated with different symbols, and with symbol size scaled to represent mass. The parameter-space that we explore in this work is shown as a grey region. The right-hand panel shows additional sensitivity regions in this observational plane for a range of other (representative) facilities. These regions are defined through combinations of spatial resolution, radial velocity and astrometric sensitivity and observational baseline (see caption for more details). Together these panels show how knowledge of the brown dwarf desert has built up to date, where our new approach contributes, and how a range of current/near-future instruments could be capable of measuring new desert discoveries. Our `search-space' is clearly a relatively unexplored separation range around low-mass M dwarfs, with discoveries having great potential for follow-up study.

We here present near-infrared spectroscopic follow-up of a subset of candidate M+UCDs identified by \citetalias{Cook2016} as M dwarfs with an increased likelihood of ultracool companionship\footnote{\citet{Cook2016} data (the NJCM catalogue) available at \url{http://vizier.cfa.harvard.edu/viz-bin/VizieR?-source=J/MNRAS/457/2192}}. In Section \ref{section:mir_excess:using_optical_spectra} we summarize the M+UCD candidates, and compare a subset of measured spectral types to our original photometric types. Section \ref{section:signatures:colour-similar} describes our spectroscopic method to confirm M+UCD candidates, and presents analysis of synthesized M+UCD systems to gauge detection confidence and assess observational requirements. We then describe our initial observations (Section \ref{section:observations}) and refinements to our method using this spectroscopy (Section \ref{section:assessing_spectral_similarity}). We apply our spectroscopic method to a preliminary set of our M+UCD candidates in Section \ref{section:results}, and discuss the results in Section \ref{section:discussion} presenting one strong M+UCD candidate and three additional M dwarfs of interest. 

Section \ref{section:conclusions} summarizes our conclusions and discusses potential future work.

\begin{figure}
\begin{center}
\includegraphics[width=.375\textwidth]{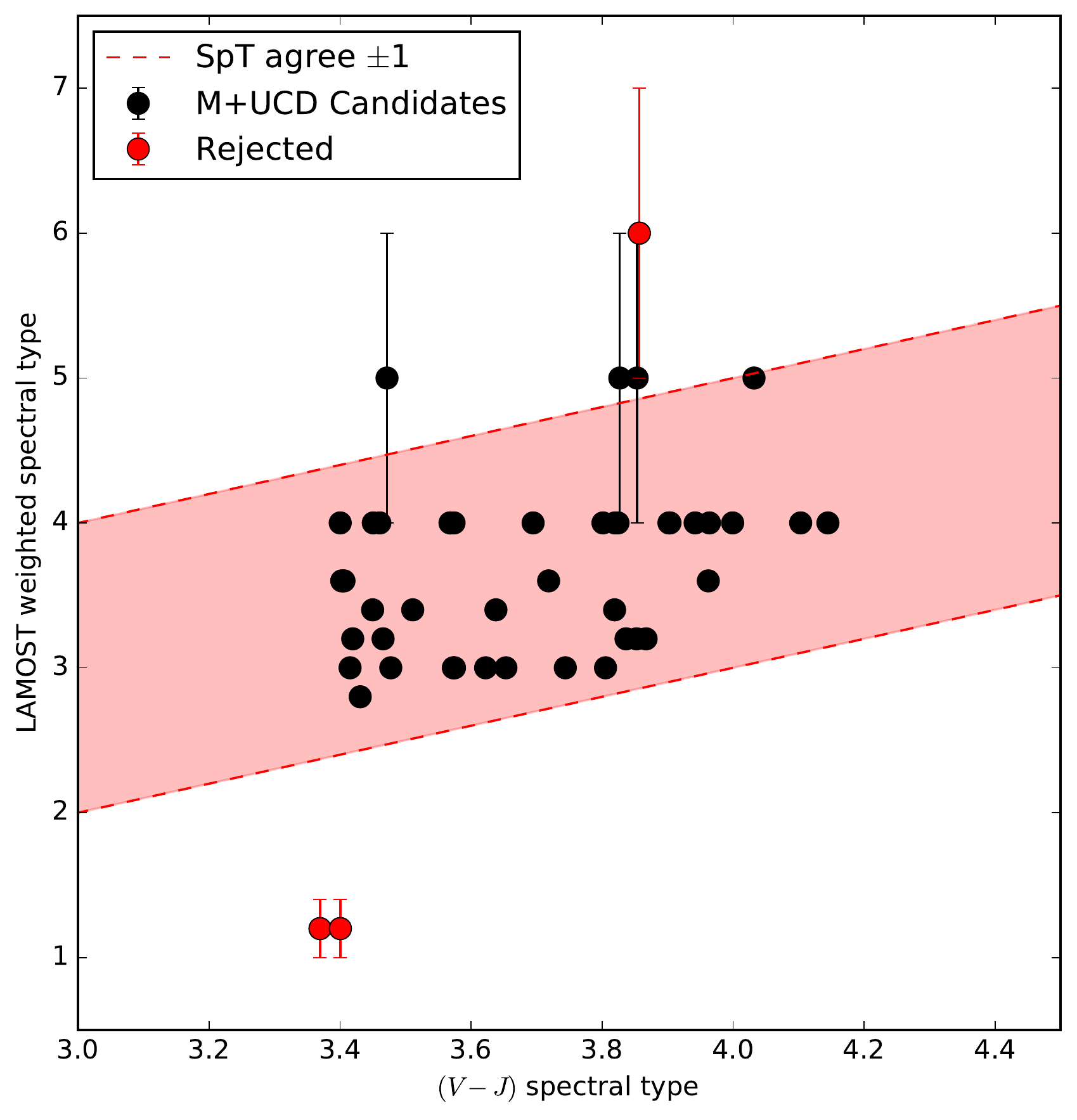}
\end{center}
\caption{Of the 46 M+UCD candidates three have photometric spectral types that differ significantly from their spectroscopic values. Uncertainties are shown for stars whose spectral types differ significantly from their photometric estimates. We consider a photometrically estimated type to be inconsistent if it differs by $\pm$1.0 from the spectroscopic value (allowing for measurement uncertainties). \label{figure:spectraltypecompare}}
\end{figure}

\section{A sample of candidate M+UCD systems}
    \label{section:mir_excess:using_optical_spectra}

    Our target sample is from the compilation of \citetalias{Cook2016}. These M dwarf candidates were selected photometrically and cleaned using strict reddening, photometric and quality constraints. Mid-infrared excesses was then assessed in the context of unresolved UCD companionship, by comparing near minus mid-infrared colours (e.g. $J-W2$) amongst subsets of optically colour-similar stars (within 0.01 mag in $g-r$, $g-i$ and $r-i$). A probability analysis then yielded M dwarfs with an increased chance ($\ge$4 times that of a random selection) of hosting an unresolved UCD companion. This process identified 1 082 M+UCD candidates, and associated colour-similar M dwarfs (to each candidate) in the catalogue. The M+UCD candidates fall into two spectral type bins with 66 per cent M3.5 and 34 per cent M4. The \JWb excess for the M3.5 candidates is $\sim$0.07 mag (equivalent to $\sim$M8--L3 companions), and for the M4 candidates it is $\sim$0.06 mag (equivalent to $\sim$L0--L4 companions).

    Poorly estimated photometric spectral types can lead to spurious M+UCD candidates. For example, underestimated types could lead to an apparent MIR excess in the absence of a UCD companion. In addition overestimated types could lead to candidates that are too bright intrinsically for measurable UCD excess signatures. To assess how beneficial it would be to have measured spectral types for our full excess sample, we have studied a sub-sample with optical spectral types measured by the {\it Large sky Area Multi-Object Fibre Spectroscopic Telescope, LAMOST} (\citealt{Cui2012,Luo2012,Zhao2012}). Provided in the LAMOST general catalogues are spectral types determined using a modified version \citep{Luo2004} of the {\sc Hammer} code \citep{Covey2014}. We combined these with spectral types following \citet{Zhong2015a}, and used a weighted mean where we had multiple spectral types (weighting by $1/\sigma^2_i$, with $\sigma_i$ the spectral type uncertainty).

    To identify M dwarfs that had inconsistent photometric spectral types (such that their M+UCD candidacy must be spurious), we measured the \VJ range of our candidate selection contours (figs 6 and 7 from \citetalias{Cook2016}) and converted these into a spectral type range using equation 12 from \citet{Lepine2013}. Objects whose true spectral types lie outside of this range are then spurious. Of the 1 082 M+UCD candidates, 46 had LAMOST spectra and thus spectral types, of which three have photometric spectral types differing significantly from their spectroscopic values (see Fig. \ref{figure:spectraltypecompare}). We predict only $\sim$7 per cent of our candidates have significantly misclassified photometric spectral types. And we thus expect a low level of spurious candidates resulting from photometric mistyping.

    As a pilot study, we chose some of the brightest M+UCD candidates from the excess sample of (\citetalias{Cook2016}), as well as a selection of associated colour-similar stars (three per candidate) that we use as control stars in our analysis method. Our selection was prioritised according to (i) the increased probability that a candidate has an unresolved UCD companion, (ii) brightness, (iii) observability and (iv) the availability of bright nearby (on-sky) colour-similar stars. Our observations of these targets will be discussed in Section \ref{section:observations}.

    \section{Simulating unresolved UCD spectroscopic signatures}
        \label{section:signatures:colour-similar}

In this section we simulate the near-infrared (NIR) spectroscopic signature of unresolved UCD companions to M3.5-4.5 dwarfs, and establish the basis for our subsequent analysis. Our general approach is to compare the NIR spectrum of an M+UCD candidate to a spectrum of a similar M dwarf that is not expected to have a UCD companion. The comparison M dwarf should have very similar optical and NIR colours to the candidate, and is referred to as a `control star'. M+UCD candidates were simulated by adding appropriately normalized spectra of an M dwarf and a UCD. Simulated control stars were based on the same M dwarf spectrum (that was used for the M+UCD candidate). However, this spectrum was multiplied through by a normalising function that was unity at 1.6 and 2.2 $\mum$, but differed by some value at 1.2 $\mum$ (we used values giving $\Delta$(J-H) of $\pm$0.01, $\pm$0.02 and $\pm$0.04 in our analysis). In practice a cubic spline fit was employed to smoothly interpolate the normalising function between these fixed values. Our approach relies on a minimum of two control stars (and ideally three) accompanying each candidate in an `observing group', so that we can compare the results of candidate to control star comparisons with those of control star to control star comparison (where the latter defines the null result).

    \subsection{Spectral ratio difference}
        \label{section:signatures:band-selection}

\begin{figure}
   \begin{center}
        \includegraphics[width=.5\textwidth]{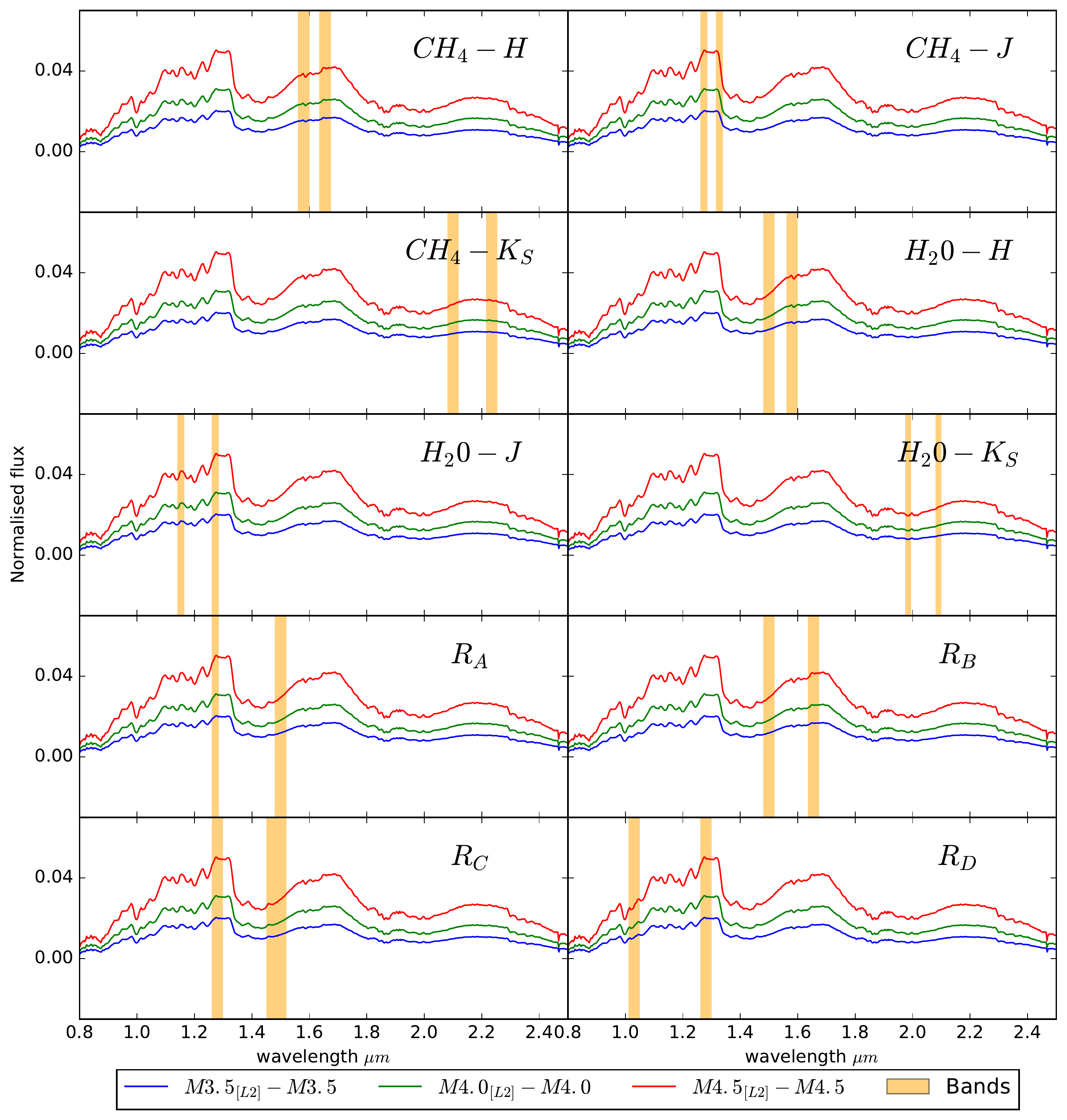}
        \end{center}
        \caption{Spectral bands from Table \ref{ch4_table_spectral_bands}. These spectral bands are compared to the subtractions for various M dwarf spectral types. The M3.5 is 2MASS J14113696+2112471 \protect\citep{Kirkpatrick2010}, the M4.5 is 2MASS J12471472-0525130 \protect\citep{Kirkpatrick2010} and the M5.5 is 2MASS J03023398-1028223 \protect\citep{Burgasser2004} and the L2 is Kelu-1 \protect\citep{Burgasser2007b}. \label{ch4_figure_spec_bands}}
\end{figure}

To provide a quantitative statistic for our spectral comparisons we based our primary comparison on spectral ratios. In the past spectral ratios have been used to identify unresolved ultracool binaries (\eg \citealt{Burgasser2010}) by assessing the spectral morphology of prominent spectral features, and comparing to typical values (for single objects). Since our approach compares candidate spectra to control star spectra (on a case-by-case basis) we instead compare a spectral ratio of a target to that of its control star, \ie we assess spectral ratio differences (Equation \ref{equation:spec-diff});

\begin{equation}
    \label{equation:spec-diff}
    \text{Spectral ratio difference} = R_1 - R_2
\end{equation}

\noindent where $R_1$ is the spectral ratio of object 1 and $R_2$ is the spectral ratio of object 2 (where we use weighted mean flux ratios). This then provides a measure of the difference in spectral morphology between a target spectrum and its control star. We also note if one normalizes both spectra (1 and 2) in the band used as the ratio denominator, the result is a measure of the flux-difference in the numerator (and can be considered as the excess flux normalized in the ratio denominator band). This means that if our numerator targets a maximum in the UCD spectrum, and the denominator targets a minimum, our spectral ratio difference will be greatest when a UCD is present in only one of the spectra. To find the optimal ratio we performed simulations using a variety of bands. The band combinations we assessed were chosen to sample some of the strong NIR absorption features in L dwarf spectra (which are also used in the spectral typing of L dwarfs; \citealt{Burgasser2010}), while avoiding regions where strong telluric absorption is an issue. In addition to the standard bands we also included two broadened bands that improve SNR (leading to the $R_{D*}$ ratio). The ratios and bands are shown in Table \ref{ch4_table_spectral_bands} and Fig. \ref{ch4_figure_spec_bands} with spectral ratios from \citep{Burgasser2010} shown for comparison.

\begin{table}
\caption{Table of spectral bands used for UCD identification via spectral ratio differences, spectral ratio difference is defined in \refequ{equation:spec-diff}. The features these spectral bands relate to can be seen in Fig. \ref{ch4_figure_spec_bands}. $^1$ From \protect\citet{Burgasser2010}. $^2$ Custom spectral bands based on those of \protect\citet{Burgasser2010} selected to optimize residual spectra while avoiding known telluric features. $^3$ After experimentation into minimizing the exposure time for observing the band was modified (see Section \ref{section:signatures:optimisingbands}). \label{ch4_table_spectral_bands}}
\begin{center}
    \begin{tabular}{lccc}
    \hline      
    Ratios   & Numerator & Denominator & Ref \\       
    \hline
    $H_20-J$ & 1.140 -- 1.165 & 1.260 -- 1.285 & 1 \\
    $CH_4-J$ & 1.315 -- 1.340 & 1.260 -- 1.285 & 1 \\ 
    $H_20-H$ & 1.480 -- 1.520 & 1.560 -- 1.600 & 1 \\
    $CH_4-H$ & 1.635 -- 1.675 & 1.560 -- 1.600 & 1 \\
    $H_20-K$ & 1.975 -- 1.995 & 2.080 -- 2.100 & 1 \\
    $CH_4-K$ & 2.215 -- 2.255 & 2.080 -- 2.120 & 1 \\
    $R_A$       & 1.260 -- 1.285 & 1.480 -- 1.520 & 2 \\
    $R_B$       & 1.635 -- 1.675 & 1.480 -- 1.520 & 2 \\
    $R_C$       & 1.260 -- 1.300 & 1.450 -- 1.520 & 2 \\
    $R_D$       & 1.260 -- 1.300 & 1.010 -- 1.050 & 2 \\
    $R_D^*$ & 1.210 --  1.350 & 0.960 --  1.100 & 3 \\
    \hline        
    \end{tabular}
    \end{center}
\end{table}

    \subsection{Boot-strapped significance}
        \label{section:signatures:Detection-threshold}

We consider an ideal observing group consisting of an M+UCD candidate plus three control stars, and analyse the spectra using a bootstrap approach. For each wavelength point in the spectra we generate a Gaussian distribution of 25 flux values (centred on the actual flux value and with a standard deviation equal to the flux uncertainty), thus creating 25 noise-variants for the candidate spectrum and 25 noise-variants for each control star spectrum. We then pair up candidate and control star spectra to yield $(25 + 1)2 \times 3$ spectral ratio difference values in the presence of an unresolved UCD companion, and pair up control star spectra to yield the same number of spectral ratio difference values in the absence of a UCD companion. These two populations of measurements are then assessed using a t-test to determine the level of significance at which they differ (see Equation \ref{equation:likelihood}).

\begin{equation}
\label{equation:likelihood}
\text{t-value} = \frac{X-Y}{\sigma_{X-Y}}
\qquad
\sigma_{X-Y} = \sqrt{\Delta X^2 + \Delta Y^2}
\end{equation}

where X is the median of the spectral ratio differences in the presence of a companion, and Y is the median in its absence.

    \subsection{Optimal ratio bands and observational requirements}
        \label{section:signatures:optimisingbands}

We calculated t-values for synthesized M+UCD candidates (according to Sections \ref{section:signatures:band-selection} and \ref{section:signatures:Detection-threshold}) using M3.5, M4.0 and M4.5 types for the primary, and L0, L2, L4 and L6 for the unresolved companion. We find overall, the $R_D$ ratio (1.26--1.3 and 1.01--1.05 $\mum$) leads to the greatest differences for such M+UCD combinations, with the greatest separation between the coloured regions (with a UCD) and the grey regions (where the UCD is absent). 

\begin{figure*}
\begin{minipage}{.8\textwidth}
        \begin{center}
        \includegraphics[trim={0 0 0 2cm}, clip, width=\textwidth]{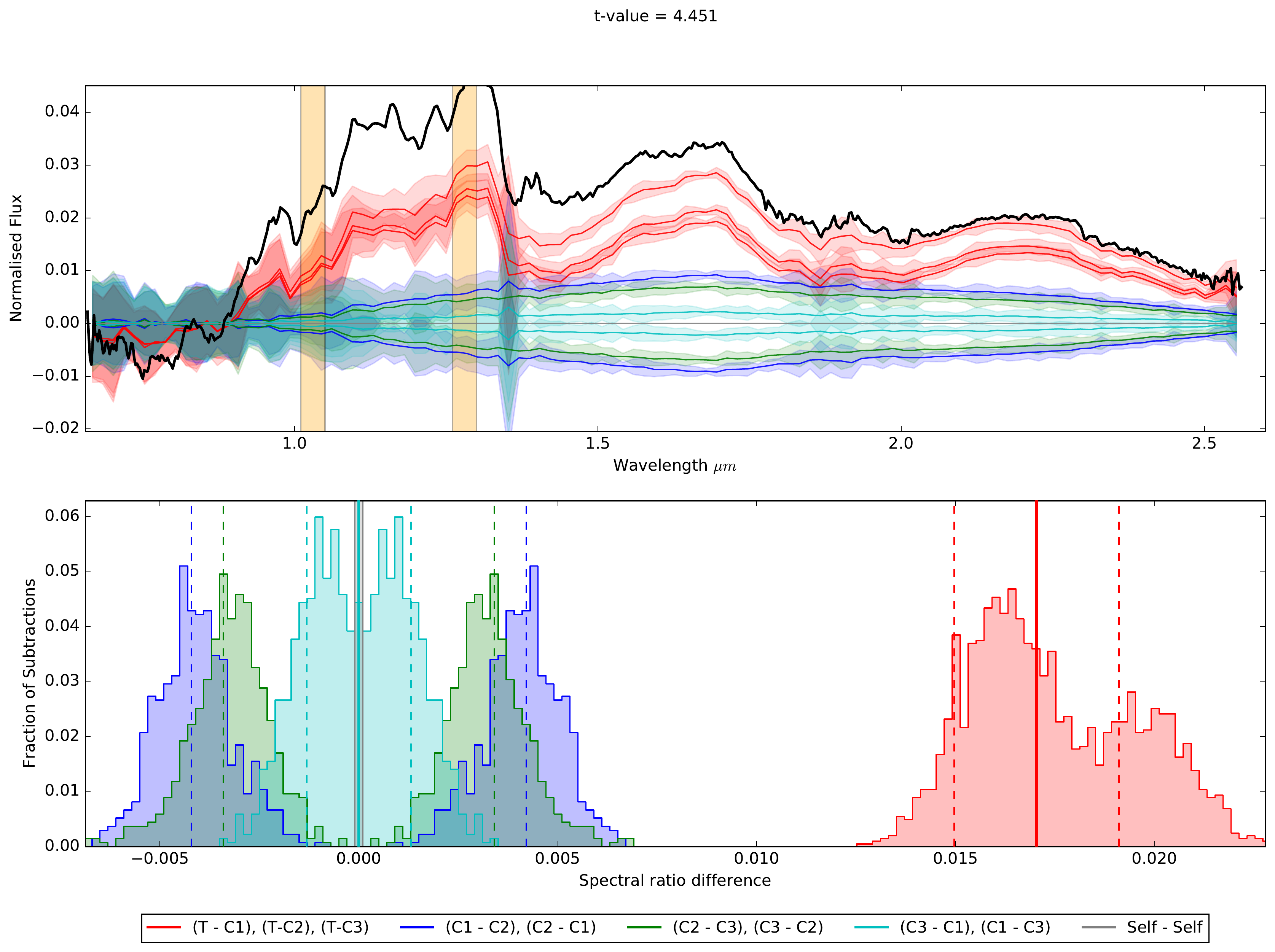}
        (a)
        \end{center}
\end{minipage}
\begin{minipage}{.8\textwidth}
        \begin{center}
        \includegraphics[trim={0 0 0 2cm}, clip, width=\textwidth]{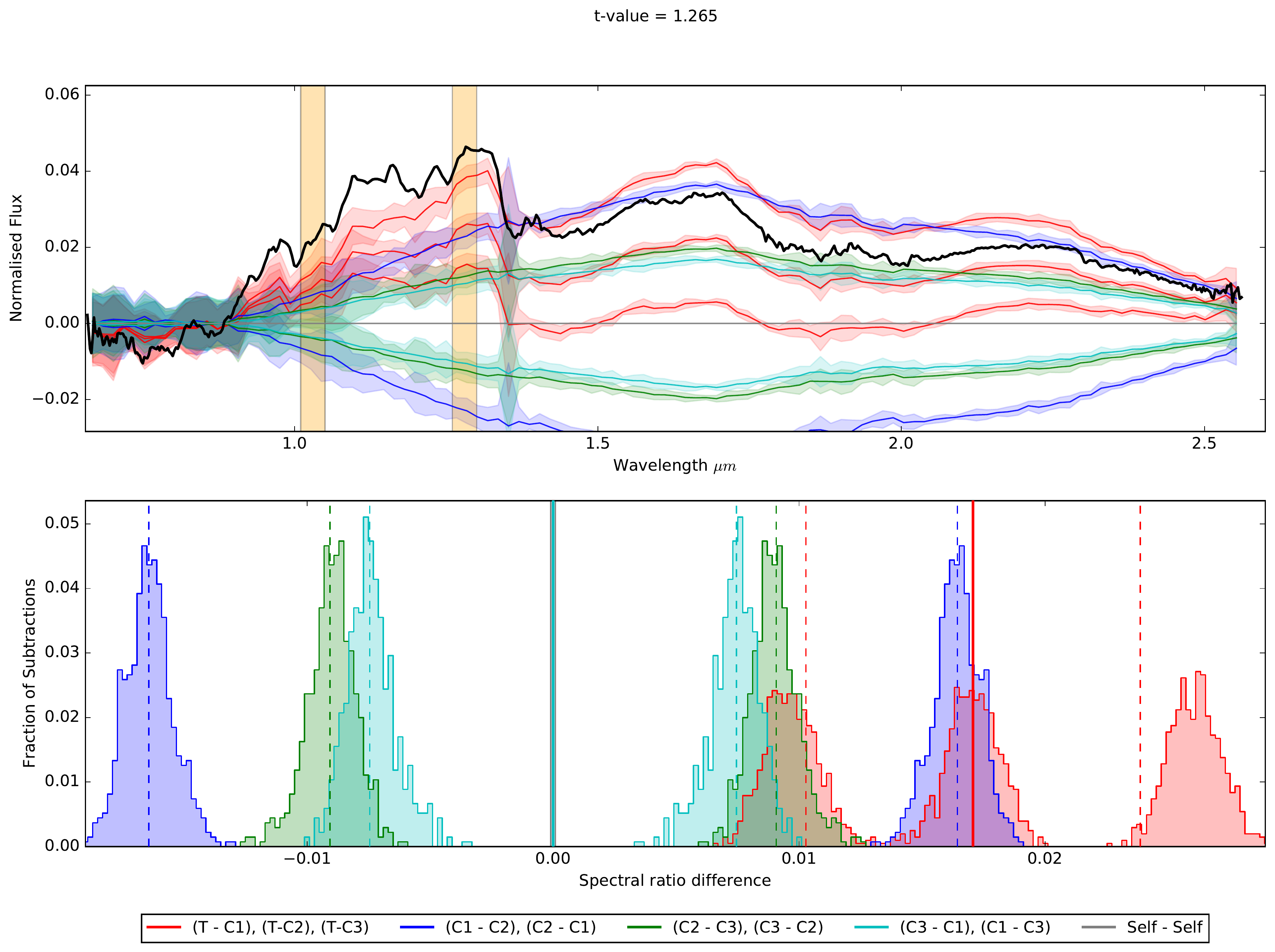}
        (b)
        \end{center}
\end{minipage}
\caption{Residual spectra plot for our simulations (a) $\Delta(J-H) = \pm 0.01$ (b) $\Delta(J-H) = \pm 0.04$. Top panels show the subtractions (target minus control subtractions $= T - CX$ and control minus control subtractions $= CX - CY$, where X and Y refer to the individual control stars). Bottom panels shows the calculated spectral ratio differences for each distribution (\refequ{equation:spec-diff}). The M dwarf used here is LP 508-14 \protect\citep{Burgasser2004} and the UCD is an L2 dwarf, Kelu-1 \protect\citep{Burgasser2007b}. Plotted in black is a comparison L0 residual (M4 is LP 508-14, \protect\citealt{Burgasser2004}, and the L0 is  2MASP J0345432+254023, \protect\citealt{BurgasserMcElwain2006}). \label{figure:specdiff_for_simulation}}
\end{figure*}

For an M4 dwarf (using the $R_D$ band) a colour-similarity of $\Delta(J-H) = \pm 0.04$ achieved a t-value of 1.3, for a $\Delta(J-H) = \pm 0.02$ the t-value was 2.2 and for $\Delta(J-H) = \pm 0.01$ the t-value was 4.6. All control stars were selected to have the lowest $\Delta(J-H)$ possible; in addition to further aid colour-similarity $\Delta(g-r)$, $\Delta(g-i)$ and $\Delta(r-i)$  were required to be less than 0.01 (as in \citetalias{Cook2016}). An example of the spectral difference results are shown in Fig. \ref{figure:specdiff_for_simulation} for $\Delta(J-H) = \pm 0.01$ and $\Delta(J-H) = \pm 0.04$. Through experimentation increasing and decreasing the bandwidth of $R_D$ we found the best t-values came from extending our bands by $\pm 0.05 \mu m$, corresponding to new spectral bands $R_D^* \equiv $ 1.21--1.35 and 0.96--1.10 $\mum$.

\begin{figure*}
\begin{minipage}{.85\textwidth}
        \begin{center}
        \includegraphics[trim={0 0 0 2cm}, clip, width=\textwidth]{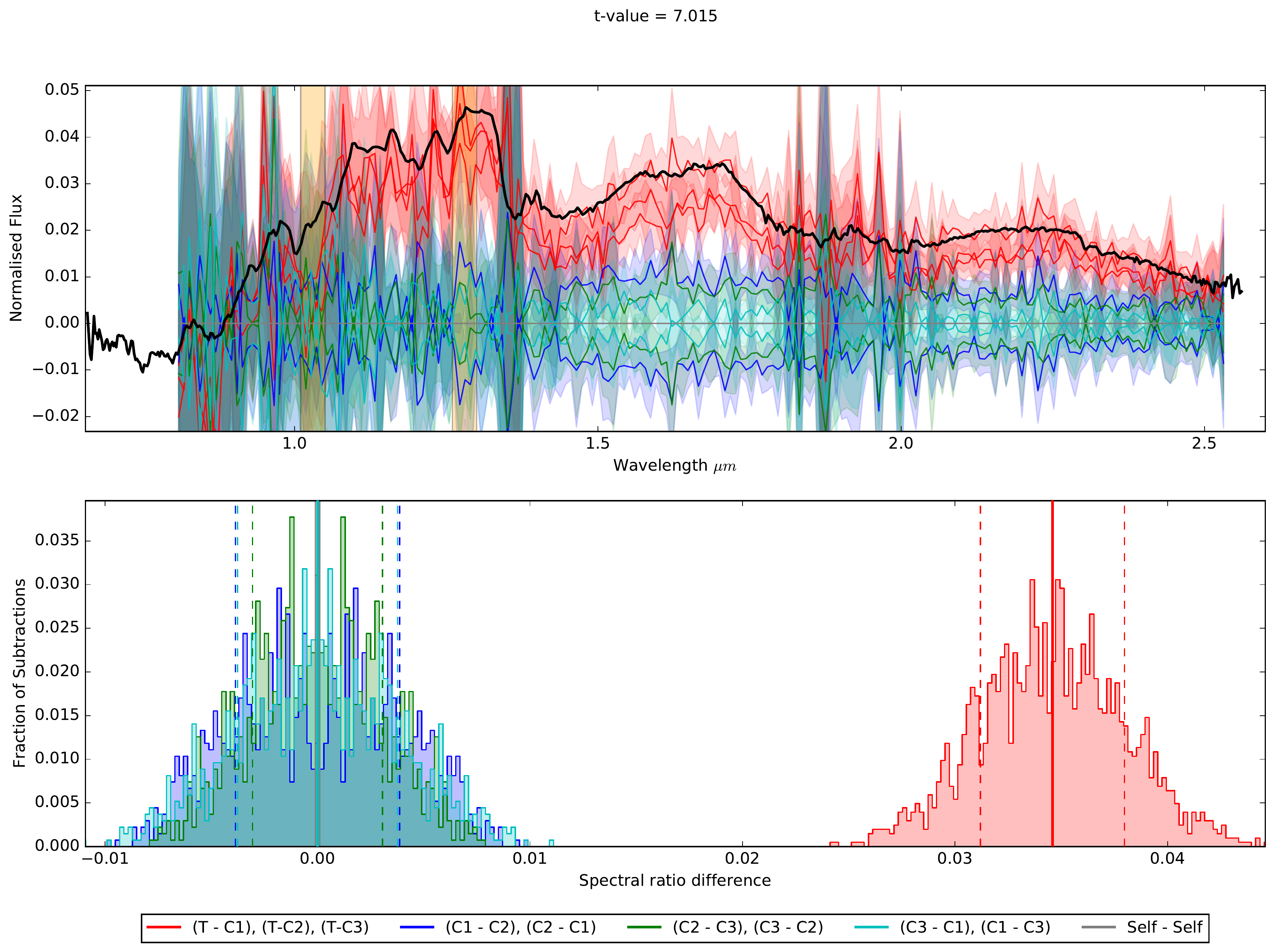}
        \\ (a)
        \end{center}
\end{minipage}
\begin{minipage}{.85\textwidth}
        \begin{center}
        \includegraphics[trim={0 0 0 2cm}, clip, width=\textwidth]{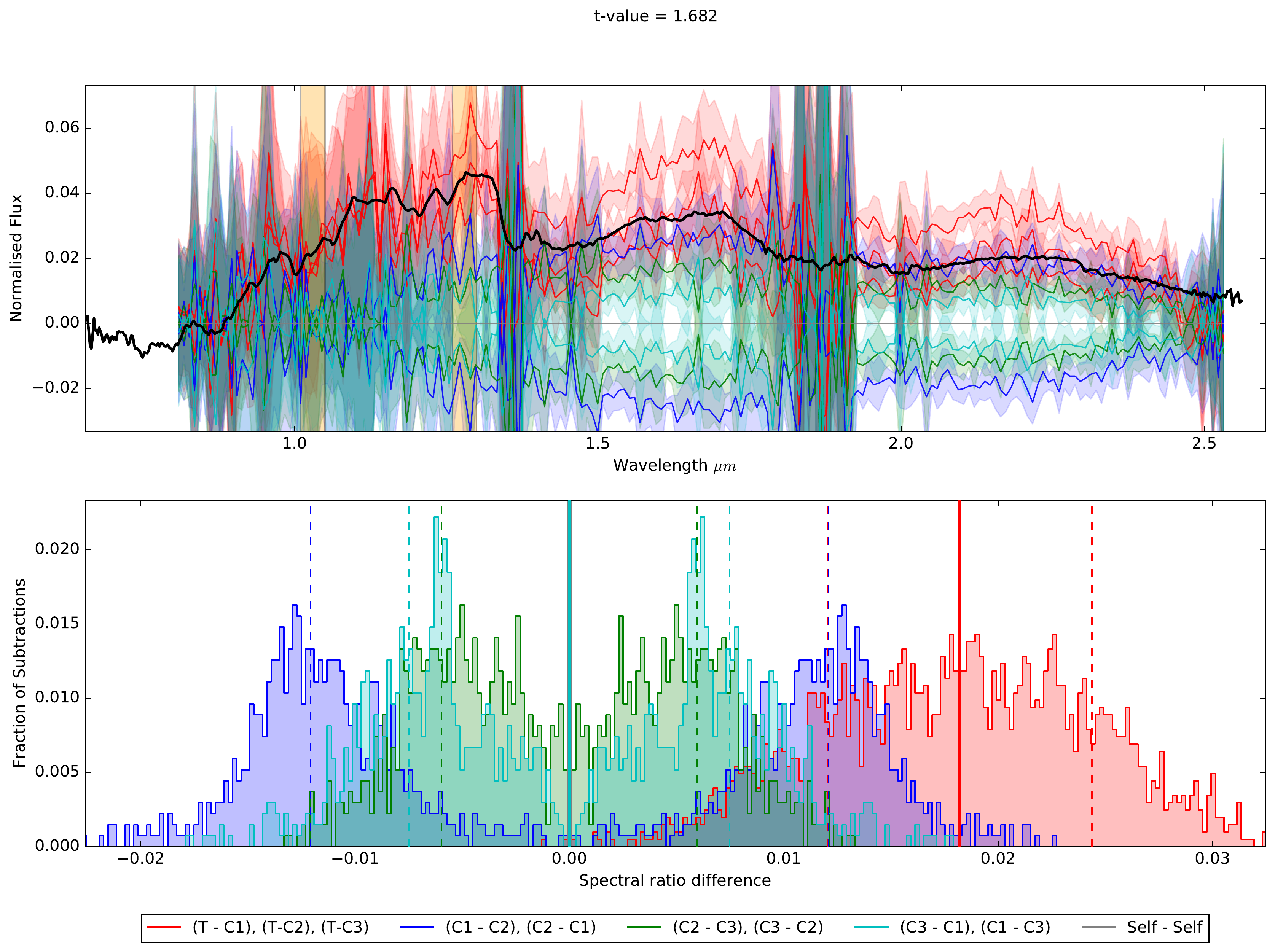}
        \\ (b)
        \end{center}
\end{minipage}
\caption{Residual spectra plot for our simulations where we simulate our observation via lowering the SNR ratio. Layout identical to Fig. \protect \ref{figure:specdiff_for_simulation}. (a) A simulated colour-similar M4 subtracted from a M4+L0, SNR = 200, $\delta \lambda / \lambda$ = 200, for $\Delta(J-H)$=0.01 (b) a simulated colour-similar M5 subtracted from a M5+L4, SNR = 200, $\delta \lambda / \lambda$ = 200, for $\Delta(J-H)$=0.02. The M dwarfs are LP 508-14 and Gliese 866AB (M4 and M5 respectively, \protect\citealt{Burgasser2004} and \protect\citealt{Burgasser2008}), the L dwarfs are 2MASP J0345432+254023 and 2MASS J21580457-1550098 (L0 and L4 respectively, \protect\citealt{Burgasser2006} and \protect\citealt{Kirkpatrick2010}). Plotted in black is a comparison L0 residual (M4 is LP 508-14, \protect\citealt{Burgasser2004}, and the L0 is  2MASP J0345432+254023, \protect\citealt{BurgasserMcElwain2006}). \label{figure:specdiff_simresults}}
\end{figure*}

\onecolumn
\LTcapwidth = 16cm

\begin{longtable}{cccccccccc}

\caption{The groups (target + controls) observed with SpeX in 2016 March. Spectral types, $SpT$, are estimated using \VJ (see \citetalias{Cook2016}).  $\alpha$ is the right acsension (WISE), $\delta$ is the declination (WISE) and $t$ is the total exposure time after combining the nods. \label{figure_observation_log}}\\
\hline
& & & & & & & & \\
WISE ID & Group & Subgroup & $\alpha$ & $\delta$ & $J$ & SpT & Date & $t$ & Airmass \\
& & & & & [mag]  & & & [min]  & \\
\hline
& & & & & & & & \\
\endfirsthead

\hline
& & & & & & & & \\
WISE ID & Group & Subgroup & $\alpha$ & $\delta$ & $J$ & SpT & Date & $t$ & Airmass \\
& & & & & [mag]  & & & [min]  & \\
\hline
& & & & & & & & \\
\endhead

J174613.19+450819.7 & 7 & Target & 17:46:13.20 & +45:08:19.8 & 13.26 & M3.5 & 2016/03/17 & 6.3 & 1.16 \\
J175142.96+425852.2 & 7 & Control 1 & 17:51:42.97 & +42:58:52.2 & 13.78 & M3.5 & 2016/03/17 & 12.5 & 1.13 \\
J172927.82+431233.5 & 7 & Control 2 & 17:29:27.82 & +43:12:33.5 & 12.79 & M3.5 & 2016/03/17 & 3.1 & 1.10 \\
J154011.95+442100.3 & 68 & Target & 15:40:11.95 & +44:21:00.3 & 13.28 & M4.0 & 2016/03/17 & 6.8 & 1.10 \\
J154933.69+423709.7 & 68 & Control 1 & 15:49:33.70 & +42:37:09.7 & 13.43 & M4.0 & 2016/03/17 & 8.6 & 1.08 \\
J160029.66+425154.3 & 68 & Control 2 & 16:00:29.67 & +42:51:54.4 & 12.98 & M4.0 & 2016/03/17 & 4.1 & 1.09 \\
J161251.50+462339.6 & 68 & Control 3 & 16:12:51.51 & +46:23:39.7 & 13.33 & M4.0 & 2016/03/17 & 7.9 & 1.12 \\
J151639.28+333630.2 & 92 & Target & 15:16:39.29 & +33:36:30.2 & 13.10 & M3.5 & 2016/03/18 & 5.3 & 1.04 \\
J150401.20+324758.6 & 92 & Control 1 & 15:04:01.20 & +32:47:58.7 & 13.84 & M3.5 & 2016/03/18 & 14.1 & 1.05 \\
J152114.93+292711.5 & 92 & Control 2 & 15:21:14.93 & +29:27:11.5 & 13.58 & M3.5 & 2016/03/18 & 10.8 & 1.05 \\
J150031.97+382736.6 & 92 & Control 3 & 15:00:31.97 & +38:27:36.7 & 13.35 & M3.5 & 2016/03/18 & 8.6 & 1.15 \\
J150642.41+324609.9 & 109 & Target & 15:06:42.41 & +32:46:10.0 & 12.39 & M3.5 & 2016/03/18 & 2.4 & 1.03 \\
J150615.82+354711.6 & 109 & Control 1 & 15:06:15.82 & +35:47:11.6 & 13.07 & M3.5 & 2016/03/18 & 5.5 & 1.04 \\
J152258.45+322504.6 & 109 & Control 2 & 15:22:58.46 & +32:25:04.7 & 12.39 & M3.5 & 2016/03/18 & 2.4 & 1.02 \\
J153743.30+324043.3 & 109 & Control 3 & 15:37:43.31 & +32:40:43.4 & 13.34 & M3.5 & 2016/03/18 & 7.9 & 1.02 \\
J144928.03+111712.9 & 124 & Target & 14:49:28.04 & +11:17:13.0 & 13.31 & M4.0 & 2016/03/19 & 6.9 & 1.05 \\
J145825.27+134738.3 & 124 & Control 1 & 14:58:25.28 & +13:47:38.4 & 13.05 & M4.0 & 2016/03/19 & 4.7 & 1.06 \\
J145830.49+171004.9 & 124 & Control 2 & 14:58:30.49 & +17:10:04.9 & 13.92 & M4.0 & 2016/03/19 & 13.9 & 1.08 \\
J151527.31+061054.6 & 124 & Control 3 & 15:15:27.31 & +06:10:54.6 & 13.72 & M4.0 & 2016/03/19 & 12.5 & 1.13 \\
J143046.74+272058.2 & 159 & Target & 14:30:46.74 & +27:20:58.2 & 13.76 & M3.5 & 2016/03/17 & 12.8 & 1.01 \\
J143927.22+265329.4 & 159 & Control 1 & 14:39:27.22 & +26:53:29.5 & 12.93 & M3.5 & 2016/03/17 & 3.3 & 1.01 \\
J141757.86+271555.8 & 159 & Control 2 & 14:17:57.86 & +27:15:55.9 & 13.72 & M3.5 & 2016/03/17 & 11.9 & 1.02 \\
J141352.49+264653.7 & 159 & Control 3 & 14:13:52.49 & +26:46:53.8 & 13.73 & M3.5 & 2016/03/17 & 11.4 & 1.03 \\
J140145.91+310640.6 & 228 & Target & 14:01:45.91 & +31:06:40.6 & 13.72 & M3.5 & 2016/03/19 & 12.4 & 1.03 \\
J141754.96+303827.4 & 228 & Control 1 & 14:17:54.97 & +30:38:27.5 & 13.52 & M3.5 & 2016/03/19 & 9.7 & 1.04 \\
J142140.35+263145.0 & 228 & Control 2 & 14:21:40.35 & +26:31:45.1 & 12.86 & M3.5 & 2016/03/19 & 3.7 & 1.02 \\
J133620.38+275852.6 & 228 & Control 3 & 13:36:20.39 & +27:58:52.7 & 13.98 & M3.5 & 2016/03/19 & 17.2 & 1.14 \\
J135939.98+271349.3 & 232 & Target & 13:59:39.98 & +27:13:49.4 & 12.79 & M3.5 & 2016/03/18 & 2.9 & 1.01 \\
J135919.47+245242.7 & 232 & Control 1 & 13:59:19.48 & +24:52:42.7 & 13.49 & M3.5 & 2016/03/18 & 9.4 & 1.00 \\
J140311.76+294227.6 & 232 & Control 2 & 14:03:11.77 & +29:42:27.6 & 13.36 & M3.5 & 2016/03/18 & 8.3 & 1.02 \\
J140922.06+320938.0 & 232 & Control 3 & 14:09:22.06 & +32:09:38.1 & 13.30 & M3.5 & 2016/03/18 & 7.3 & 1.03 \\
J133709.98+051838.0 & 282 & Target & 13:37:09.99 & +05:18:38.0 & 13.60 & M3.5 & 2016/03/18 & 9.7 & 1.05 \\
J135037.01+052648.3 & 282 & Control 1 & 13:50:37.02 & +05:26:48.4 & 13.69 & M3.5 & 2016/03/18 & 11.3 & 1.05 \\
J135218.99+065447.2 & 282 & Control 2 & 13:52:19.00 & +06:54:47.2 & 12.93 & M3.5 & 2016/03/18 & 3.3 & 1.03 \\
J135342.62+030317.3 & 282 & Control 3 & 13:53:42.62 & +03:03:17.3 & 12.89 & M3.5 & 2016/03/18 & 3.7 & 1.02 \\
J131246.68+301857.5 & 340 & Target & 13:12:46.68 & +30:18:57.6 & 13.65 & M4.0 & 2016/03/19 & 11.1 & 1.02 \\
J133526.02+291402.1 & 340 & Control 1 & 13:35:26.03 & +29:14:02.1 & 13.85 & M4.0 & 2016/03/19 & 14.3 & 1.01 \\
J134722.80+314804.7 & 340 & Control 2 & 13:47:22.80 & +31:48:04.7 & 13.68 & M4.0 & 2016/03/19 & 11.4 & 1.03 \\
J132515.09+224902.4 & 340 & Control 3 & 13:25:15.09 & +22:49:02.4 & 13.14 & M4.0 & 2016/03/19 & 5.9 & 1.04 \\
J130340.78+152551.9 & 360 & Target & 13:03:40.78 & +15:25:51.9 & 12.78 & M3.5 & 2016/03/17 & 2.9 & 1.00 \\
J132523.20+174000.9 & 360 & Control 1 & 13:25:23.21 & +17:40:01.0 & 13.62 & M3.5 & 2016/03/17 & 10.7 & 1.00 \\
J124919.38+210618.2 & 360 & Control 2 & 12:49:19.39 & +21:06:18.3 & 12.85 & M3.5 & 2016/03/17 & 3.3 & 1.03 \\
J131145.84+084345.0 & 360 & Control 3 & 13:11:45.84 & +08:43:45.1 & 13.66 & M3.5 & 2016/03/17 & 11.4 & 1.04 \\
J122352.96+052659.9 & 466 & Target & 12:23:52.97 & +05:26:59.9 & 12.96 & M3.5 & 2016/03/17 & 3.8 & 1.04 \\
J122932.96+075423.5 & 466 & Control 1 & 12:29:32.96 & +07:54:23.6 & 13.14 & M3.5 & 2016/03/17 & 5.3 & 1.02 \\
J120916.25+051754.2 & 466 & Control 2 & 12:09:16.25 & +05:17:54.3 & 13.67 & M3.5 & 2016/03/17 & 10.1 & 1.02 \\
J125055.41+043050.3 & 466 & Control 3 & 12:50:55.41 & +04:30:50.3 & 13.96 & M3.5 & 2016/03/17 & 16.5 & 1.04 \\
J122043.33+203120.6 & 476 & Target & 12:20:43.34 & +20:31:20.7 & 13.32 & M4.0 & 2016/03/19 & 7.1 & 1.00 \\
J124740.40+233615.7 & 476 & Control 1 & 12:47:40.41 & +23:36:15.7 & 12.73 & M3.5 & 2016/03/19 & 3.1 & 1.00 \\
J121822.55+285654.2 & 476 & Control 2 & 12:18:22.55 & +28:56:54.3 & 12.79 & M3.5 & 2016/03/19 & 3.1 & 1.02 \\
J114605.14+234707.5 & 476 & Control 3 & 11:46:05.15 & +23:47:07.6 & 14.01 & M4.0 & 2016/03/19 & 16.9 & 1.05 \\
J114857.72+073046.3 & 550 & Target & 11:48:57.73 & +07:30:46.3 & 12.68 & M4.0 & 2016/03/18 & 2.7 & 1.02 \\
J122213.43+091128.9 & 550 & Control 1 & 12:22:13.44 & +09:11:29.0 & 13.64 & M4.0 & 2016/03/18 & 9.9 & 1.02 \\
J122200.86+121753.1 & 550 & Control 2 & 12:22:00.86 & +12:17:53.1 & 13.13 & M4.0 & 2016/03/18 & 5.7 & 1.01 \\
J115522.06+002657.3 & 550 & Control 3 & 11:55:22.07 & +00:26:57.3 & 12.03 & M4.0 & 2016/03/18 & 2.1 & 1.09 \\
J104507.41+181311.0 & 697 & Target & 10:45:07.41 & +18:13:11.1 & 12.36 & M3.5 & 2016/03/18 & 2.3 & 1.01 \\
J104540.21+174228.0 & 697 & Control 1 & 10:45:40.22 & +17:42:28.0 & 12.46 & M3.5 & 2016/03/18 & 2.5 & 1.01 \\
J110355.23+153411.7 & 697 & Control 2 & 11:03:55.23 & +15:34:11.7 & 12.00 & M3.5 & 2016/03/18 & 2.1 & 1.02 \\
J111056.11+180251.5 & 697 & Control 3 & 11:10:56.12 & +18:02:51.5 & 13.28 & M3.5 & 2016/03/18 & 7.5 & 1.01 \\
J102239.45+053345.5 & 751 & Target & 10:22:39.46 & +05:33:45.6 & 13.13 & M3.5 & 2016/03/17 & 5.7 & 1.04 \\
J103128.01+054011.4 & 751 & Control 1 & 10:31:28.01 & +05:40:11.4 & 14.11 & M3.0 & 2016/03/17 & 18.9 & 1.04 \\
J100449.01+135334.6 & 751 & Control 2 & 10:04:49.01 & +13:53:34.7 & 13.55 & M3.5 & 2016/03/17 & 10.1 & 1.09 \\
J095202.99+020820.5 & 751 & Control 3 & 09:52:03.00 & +02:08:20.6 & 12.66 & M3.5 & 2016/03/17 & 2.8 & 1.22 \\
J102115.33+422822.5 & 757 & Target & 10:21:15.33 & +42:28:22.5 & 13.35 & M3.5 & 2016/03/19 & 7.7 & 1.10 \\
J101655.64+415752.6 & 757 & Control 1 & 10:16:55.65 & +41:57:52.6 & 13.00 & M3.5 & 2016/03/19 & 4.1 & 1.12 \\
J102402.03+374148.4 & 757 & Control 2 & 10:24:02.04 & +37:41:48.4 & 13.76 & M3.5 & 2016/03/19 & 12.7 & 1.11 \\
J103056.60+471237.9 & 757 & Control 3 & 10:30:56.60 & +47:12:38.0 & 13.55 & M3.5 & 2016/03/19 & 10.0 & 1.21 \\
J102051.15+474023.9 & 761 & Target & 10:20:51.16 & +47:40:24.0 & 13.15 & M3.5 & 2016/03/19 & 6.1 & 1.14 \\
J103540.64+472827.4 & 761 & Control 1 & 10:35:40.64 & +47:28:27.4 & 13.99 & M3.5 & 2016/03/19 & 17.1 & 1.14 \\
J103236.96+455934.3 & 761 & Control 2 & 10:32:36.96 & +45:59:34.4 & 13.87 & M3.5 & 2016/03/19 & 15.2 & 1.11 \\
J095850.08+451018.8 & 761 & Control 3 & 09:58:50.08 & +45:10:18.8 & 12.89 & M3.5 & 2016/03/19 & 3.7 & 1.13 \\
J100202.50+074136.3 & 800 & Target & 10:02:02.51 & +07:41:36.4 & 12.53 & M3.5 & 2016/03/18 & 2.5 & 1.03 \\
J100211.65+075540.6 & 800 & Control 1 & 10:02:11.65 & +07:55:40.6 & 13.40 & M3.5 & 2016/03/18 & 8.3 & 1.03 \\
J100515.21+110551.5 & 800 & Control 2 & 10:05:15.21 & +11:05:51.6 & 13.37 & M3.5 & 2016/03/18 & 8.4 & 1.03 \\
J101911.78+101143.2 & 800 & Control 3 & 10:19:11.79 & +10:11:43.3 & 12.59 & M3.5 & 2016/03/18 & 2.6 & 1.03 \\
J093819.44+565237.6 & 862 & Target & 09:38:19.45 & +56:52:37.7 & 12.97 & M4.0 & 2016/03/19 & 4.1 & 1.26 \\
J092416.41+555952.1 & 862 & Control 1 & 09:24:16.41 & +55:59:52.2 & 13.16 & M4.0 & 2016/03/19 & 5.6 & 1.24 \\
J090223.85+620747.4 & 862 & Control 2 & 09:02:23.85 & +62:07:47.4 & 13.54 & M4.0 & 2016/03/19 & 9.7 & 1.35 \\
J092142.11+643630.5 & 862 & Control 3 & 09:21:42.12 & +64:36:30.6 & 12.12 & M4.0 & 2016/03/19 & 2.1 & 1.49 \\
J092547.70+430605.3 & 890 & Target & 09:25:47.71 & +43:06:05.3 & 12.59 & M4.0 & 2016/03/19 & 2.5 & 1.14 \\
J100815.76+420546.1 & 890 & Control 1 & 10:08:15.77 & +42:05:46.1 & 13.49 & M4.0 & 2016/03/19 & 8.9 & 1.20 \\
J092438.33+383415.6 & 890 & Control 2 & 09:24:38.33 & +38:34:15.7 & 12.72 & M4.0 & 2016/03/19 & 2.9 & 1.08 \\
J091551.28+470403.7 & 890 & Control 3 & 09:15:51.29 & +47:04:03.7 & 12.94 & M4.0 & 2016/03/19 & 3.9 & 1.13 \\
J090908.53+354727.5 & 918 & Target & 09:09:08.53 & +35:47:27.6 & 13.20 & M3.5 & 2016/03/18 & 6.1 & 1.04 \\
J090520.58+324153.3 & 918 & Control 1 & 09:05:20.58 & +32:41:53.4 & 14.05 & M3.5 & 2016/03/18 & 18.7 & 1.03 \\
J090322.79+394915.2 & 918 & Control 2 & 09:03:22.79 & +39:49:15.2 & 13.41 & M3.5 & 2016/03/18 & 8.7 & 1.08 \\
J093315.96+355255.4 & 918 & Control 3 & 09:33:15.97 & +35:52:55.5 & 13.38 & M3.5 & 2016/03/18 & 8.2 & 1.05 \\
J090114.01+331945.1 & 927 & Target & 09:01:14.01 & +33:19:45.2 & 13.18 & M4.0 & 2016/03/18 & 5.4 & 1.08 \\
J092037.17+363745.6 & 927 & Control 1 & 09:20:37.17 & +36:37:45.7 & 13.96 & M4.0 & 2016/03/18 & 17.3 & 1.10 \\
J084240.19+262513.0 & 927 & Control 2 & 08:42:40.19 & +26:25:13.0 & 13.60 & M4.0 & 2016/03/18 & 10.1 & 1.01 \\
J093334.33+381013.3 & 927 & Control 3 & 09:33:34.33 & +38:10:13.3 & 13.44 & M4.0 & 2016/03/18 & 9.1 & 1.07 \\
J085410.72+443149.3 & 942 & Target & 08:54:10.72 & +44:31:49.4 & 12.92 & M3.5 & 2016/03/18 & 3.5 & 1.27 \\
J084026.52+435854.7 & 942 & Control 1 & 08:40:26.53 & +43:58:54.8 & 13.74 & M3.5 & 2016/03/18 & 13.7 & 1.21 \\
J090549.19+482615.8 & 942 & Control 2 & 09:05:49.19 & +48:26:15.9 & 13.28 & M3.5 & 2016/03/18 & 6.6 & 1.24 \\
J080946.15+464349.0 & 942 & Control 3 & 08:09:46.16 & +46:43:49.1 & 13.38 & M3.5 & 2016/03/18 & 9.2 & 1.13 \\
J085237.84+431441.7 & 946 & Target & 08:52:37.84 & +43:14:41.8 & 13.42 & M4.0 & 2016/03/17 & 8.1 & 1.11 \\
J092250.12+432738.6 & 946 & Control 1 & 09:22:50.12 & +43:27:38.7 & 12.97 & M4.0 & 2016/03/17 & 4.3 & 1.10 \\
J091453.37+443448.5 & 946 & Control 2 & 09:14:53.37 & +44:34:48.6 & 13.54 & M4.0 & 2016/03/17 & 9.7 & 1.13 \\
J090309.84+354215.1 & 946 & Control 3 & 09:03:09.84 & +35:42:15.1 & 13.51 & M4.0 & 2016/03/17 & 9.2 & 1.10 \\
J084819.77+430919.4 & 956 & Target & 08:48:19.77 & +43:09:19.5 & 12.91 & M3.5 & 2016/03/17 & 3.6 & 1.09 \\
J091529.41+430447.0 & 956 & Control 1 & 09:15:29.42 & +43:04:47.0 & 13.82 & M3.5 & 2016/03/17 & 12.5 & 1.10 \\
J090408.84+383249.0 & 956 & Control 2 & 09:04:08.85 & +38:32:49.1 & 13.22 & M3.5 & 2016/03/17 & 6.9 & 1.05 \\
J085431.35+374703.4 & 956 & Control 3 & 08:54:31.35 & +37:47:03.4 & 13.40 & M3.5 & 2016/03/17 & 9.0 & 1.06 \\
J084530.09+192606.6 & 961 & Target & 08:45:30.09 & +19:26:06.7 & 13.79 & M3.5 & 2016/03/19 & 12.7 & 1.11 \\
J084634.84+191526.0 & 961 & Control 1 & 08:46:34.84 & +19:15:26.0 & 13.75 & M3.5 & 2016/03/19 & 11.9 & 1.09 \\
J084536.83+183555.3 & 961 & Control 2 & 08:45:36.83 & +18:35:55.3 & 12.99 & M3.5 & 2016/03/19 & 3.7 & 1.05 \\
J084800.51+212638.8 & 961 & Control 3 & 08:48:00.52 & +21:26:38.9 & 13.89 & M3.5 & 2016/03/19 & 14.7 & 1.02 \\
J082443.15+044240.8 & 985 & Target & 08:24:43.16 & +04:42:40.9 & 13.33 & M3.5 & 2016/03/17 & 7.3 & 1.12 \\
J081557.78+005921.6 & 985 & Control 1 & 08:15:57.79 & +00:59:21.6 & 13.71 & M3.5 & 2016/03/17 & 10.7 & 1.08 \\
J084042.24+081202.4 & 985 & Control 2 & 08:40:42.25 & +08:12:02.4 & 12.65 & M3.5 & 2016/03/17 & 2.6 & 1.07 \\
J084338.17+080419.0 & 985 & Control 3 & 08:43:38.18 & +08:04:19.1 & 13.70 & M3.5 & 2016/03/17 & 11.5 & 1.04 \\
\hline
\end{longtable}
\twocolumn

\begin{figure*}
    \begin{center}
    \begin{minipage}{.4825\textwidth}
    \begin{center}
    \includegraphics[width=1.05\textwidth,trim={0 0 0 1.5cm},clip] {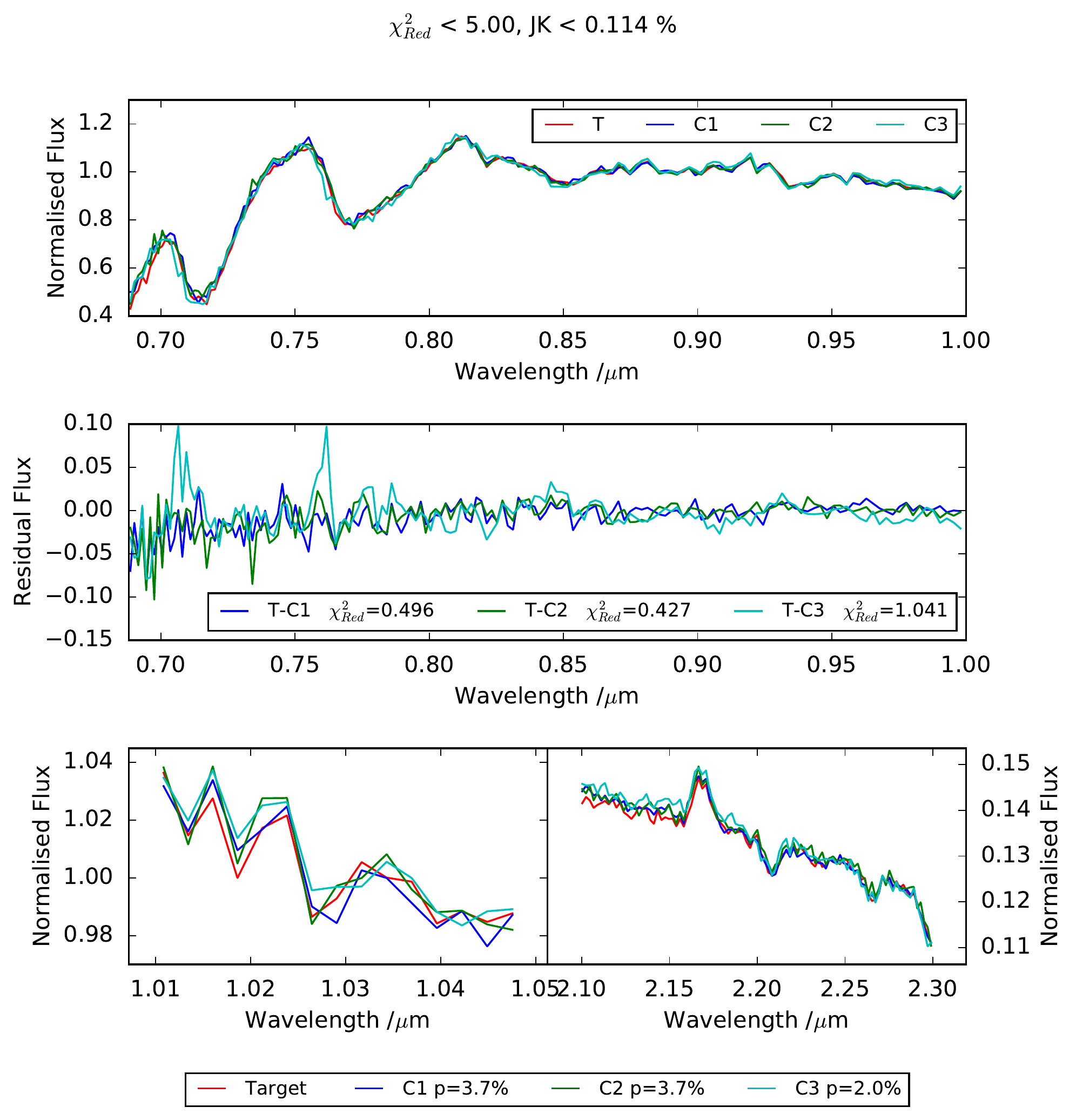}
    \\ (a)
    \end{center}
    \end{minipage}\qquad
    \begin{minipage}{.4825\textwidth}
    \begin{center}
    \includegraphics[width=1.05\textwidth,trim={0 0 0 1.5cm},clip] {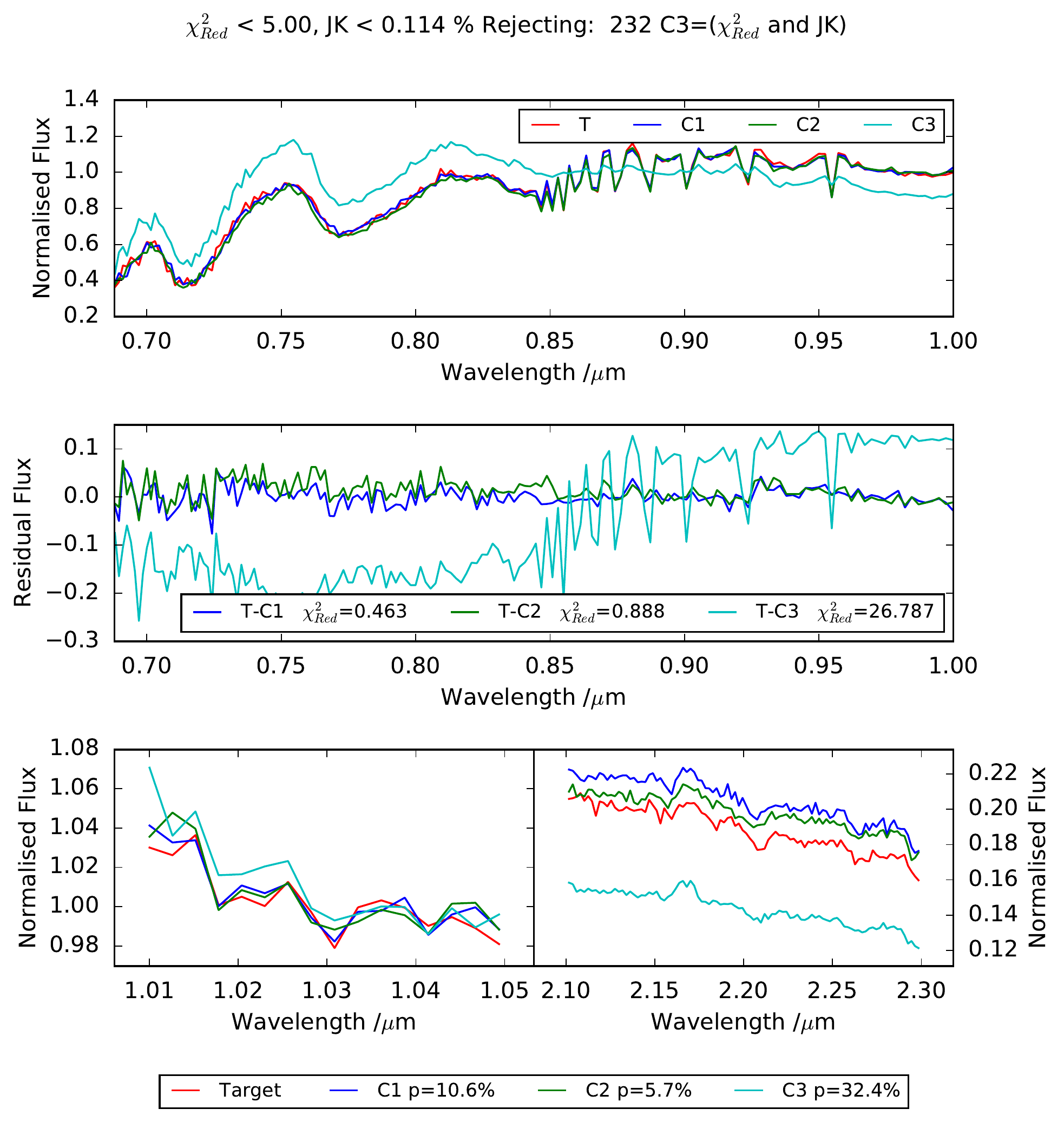}
    \\ (b)
    \end{center}
    \end{minipage}
    \end{center}
    \caption{Example of (a) a clean group (where no control stars were rejected), and (b) a group where one control star is rejected. Top panel of each plot shows the optical part of the SpeX spectrum, middle panel shows the residuals between the target and control stars, bottom panels show the numerator and denominator region of the spectrum used to calculate the \JKcut band spectral ratio for both target and control stars, where we normalize the mean flux of the numerator to unity. The simulated cases use the same spectra as Fig. \protect\ref{figure:specdiff_for_simulation}. T here is the target (M+UCD candidate), and C1, C2 and C3 are the control stars (for each observation group).\label{paper2_figure_control_rejction}}
\end{figure*}

\noindent To provide constraints on observational requirements, as well as information about the parameter-space of detectable companions, we investigated expected changes in our t-value results due to differing SNR and resolution ($\delta \lambda / \lambda$ (as well as change in primary and secondary spectral type). We ran a series of tests designed to identify the parameters that achieve (i) an optimal result, with t-values in the range 3--5, as well as (ii) a minimal result, with t-values in the range 2--3, using the shortest possible telescope exposure time (\ie combination of low SNR and low resolution). We thus require spectroscopy with a SNR of at least 125 and  $\delta \lambda / \lambda > 25$. For an optimal result we require spectroscopy with a SNR $\sim$200 and $\delta \lambda / \lambda \sim 200$. Increasing the SNR can be achieved by reducing the resolution (via binning up the pixels and applying a Gaussian smoothing function).

    \subsection{Residual spectra}
        \label{section:signatures:sim-results}

Our spectroscopic difference ratios are indicative of the flux excess in the M+UCD candidates (relative to the control stars, see Section \ref{section:signatures:band-selection}) when normalized in the denominator band. To obtain a more detailed view of this excess flux we also plot residual spectra, resulting from both target minus control subtractions and control minus control subtractions. The target minus control subtraction residuals should show a trace of the unresolved UCD spectrum, and the control minus control subtraction residuals should indicate the level of residual excess one can expect for the null case.

Fig. \ref{figure:specdiff_for_simulation} shows the residuals for our simulated spectra (top panels) with target minus control in red, and for control minus control in blue/green/cyan. For comparison we also plot simulated results for a brighter L0 UCD companion and an ideal case ($\Delta(J-H)=0$) control star, as a black line.

The bottom panels show histograms of the spectral ratio difference values for the relevant combinations of simulated spectra. The greater the separation between the target minus control distribution and the control minus control distribution, the higher the t-value will be and the more significant the UCD detection. Fig. \ref{figure:specdiff_simresults} shows simulations results at reduced SNR (according to Section \ref{section:signatures:optimisingbands}).

    \section{Observation and data reduction}
        \label{section:observations}

We obtained low-resolution ($\lambda/\Delta\lambda \sim150$), near-infrared spectra from SpeX on NASA's {\itshape Infrared Telescope Facility} \citep[IRTF, ][]{Rayner2003} using the 0.5$\times$15\arcsec slit (post-upgrade PRISM mode, $\sim$0.7 - 2.52 $\mu$m) on 2016 <March 17, 18 and 19 (2016A051, mean seeing of 0.84, 0.85 and 0.70, respectively). 

Observations were obtained of 28 M+UCD candidate M dwarfs, and for most of these we also targeted three colour-similar control stars per candidate (whose optical SDSS colours are within 0.01 mag of the candidate; see Section \ref{section:mir_excess:using_optical_spectra}) which were reasonably close in airmass. In one case we were only able to observe two control stars to accompany the candidate (due to time constraints), leading to a total of 83 control stars being observed (in a standard ABBA fashion). Exposure times\footnote{Exposure times were calculated using the web-based input form for SpeX \url{http://irtfweb.ifa.hawaii.edu/cgi-bin/spex/spex\_calc2.cgi}} were calculated to give an SNR greater than $\sim150$ at $1.05 \mum$.

We observed each group of M dwarfs (M+UCD candidate plus control stars) consecutively to ensure observing conditions were as similar as possible and also observed one standard star (A0V-type star or similar) close in time and airmass. Flat-fields and argon lamp calibrations were obtained to accompany each group.

The data were reduced using the facility-provided {\sc SpeXTools} package \citep{Cushing2004} that automatically subtracts all AB nods, extracts the spectrum, flat-fields and wavelength calibrates the spectra. We corrected for telluric absorption using the {\sc xtellcor} program \citep{Vacca2003}. Finally spectra were binned up by a factor of 5 to further increase the SNR. Our spectroscopic observations are summarized in Table \ref{figure_observation_log}.

    \section{Identifying spectrally similar control stars}
        \label{section:assessing_spectral_similarity}

Our pre-observation selection of control stars was based on colour-similarity (with associated targets) using available SDSS photometry (see Section \ref{section:mir_excess:using_optical_spectra}). With observed spectra in-hand we also carried out spectroscopic analysis to further improve on this similarity assessment.

We used a reduced chi-squared analysis to compare the optical region of each target with its associated control stars (in the $<$ 1 $\mum$ range, where a UCD has little-to-no contribution to the flux). As a second condition we required each target and its associated control stars have similar \JKcut flux ratios. We define our \JKcut ratio using wavelength bands 1.01--1.05 $\mum$ and 2.10--2.30 \mum. Within these bands we expect the flux contribution from a UCD to be relatively low (compared to the wavelength region between these bands).

Control stars were rejected if their spectroscopic difference in the optical (compared to their associated target) amounted to $\chi^{2}_{\rm Red} > 5$. We also rejected control stars whose \JKcut ratio was more than 11.4 per cent different to their associated target, which represents the 2$\sigma$ range for the full control star sample. Fig. \ref{paper2_figure_control_rejction} illustrates our rejection procedure for two example groups. In Fig. \ref{paper2_figure_control_rejction}a, no control stars were rejected from the group, and in Fig. \ref{paper2_figure_control_rejction}b, a single control star was rejected as a result of failing both the optical reduced chi-squared condition and the near-infrared \JKcut requirement.

Of the 28 observed groups, there were eight groups for which one control star was rejected, and three where two or more control stars were rejected (see Table \ref{table-rejection_table}). Groups with two or more rejected control stars were removed from further analysis. Thus, 25 groups with two or three colour and spectroscopically similar control stars were taken forward for further analysis.

\section{Candidate and control star analysis}
        \label{section:results}

For each observation group we measured spectral ratio differences (see Section \ref{section:signatures:band-selection}), t-values (see Section \ref{section:signatures:Detection-threshold}), and spectroscopic residuals (see Section \ref{section:signatures:sim-results}). Our t-value calculations were made using three different approaches so as to put our final results in a useful context.

\begin{enumerate}[
label=\arabic*)
]

\item M+UCD candidate minus control \\

Spectral ratio differences and t-values were calculated for each group using the M+UCD candidate and control star pairings, following the same bootstrap approach described in Section \ref{section:signatures:Detection-threshold}. This provided a measure for the strength of any unresolved UCD companions around the candidates, and was carried out for the 25 groups with the required number of control stars. \\

\item Control star minus control star\\

For each group one of the control stars was treated in an identical manner to an M+UCD candidate (in the type 1 analysis), with spectral ratio differences and t-values calculated accordingly. Since the control stars are defined as having no detectable \JWb excess, any near-infrared excess would presumably come from a non-UCD origin. This therefore allowed us to assess false positives with our data set. We could only carry out this analysis for groups that had three usable control stars, which amounted to 16 of the 25 groups.

\noindent Thus we had 48 (16$\times$3) combinations of subtractions (\ie control 1 compared to control 2 and control 3, control 2 compared to control 1 and control 3, etc.) on which to base our false positive count. \\

\item Model minus control \\

For each group we selected one of the control stars (at random) and added to its spectrum a known L2 UCD (Kelu-1, \citealt{Burgasser2007b}, flux-normalized appropriately). Spectral ratio differences and t-values were then calculated by treating these artificially generated M+UCD objects in the same way as the candidates (in the type 1 analysis), pairing them up with un-altered control stars from the group. This therefore allowed us to assess the expected results for the case where every analysed group had an L2 signature injected into the spectrum of its `candidate M dwarf'. This therefore provides an assessment of how effectively the colour-similar control stars allow a known L2 signature to be uncovered through our analysis. As for the type 1 analysis we could only carry this out for the 25 groups that had two or three usable control stars.

\end{enumerate}

\noindent We chose to use a t-value of 1.75 to indicate possible detections. This is close to our minimal requirement and represented a good trade-off between identifying interesting candidates and minimizing false positives. Table \ref{ch4_table_tvalues_all} presents our t-values for all 25 candidate M+UCD systems considered here.

    \section{Results and discussion}
        \label{section:discussion}

\begin{figure*}
\begin{minipage}{.85\textwidth}
        \begin{center}
        \includegraphics[trim={0 0 0 2cm}, clip, width=\textwidth]{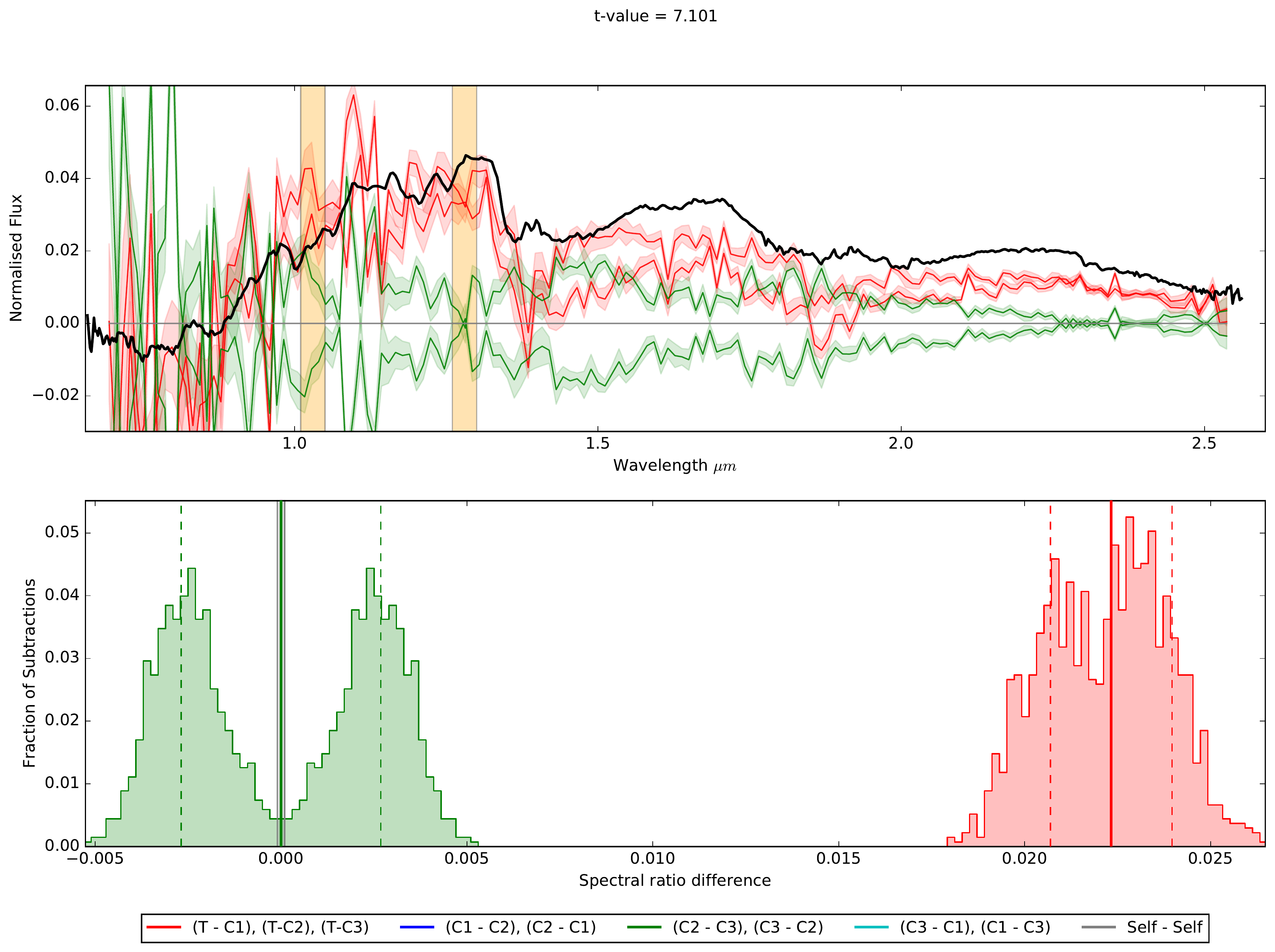}
        \\ (a) WISE J100202.50+074136.3, part of group 800.
        \end{center}
\end{minipage}
\begin{minipage}{.85\textwidth}
        \begin{center}
        \includegraphics[trim={0 0 0 2cm}, clip, width=\textwidth]{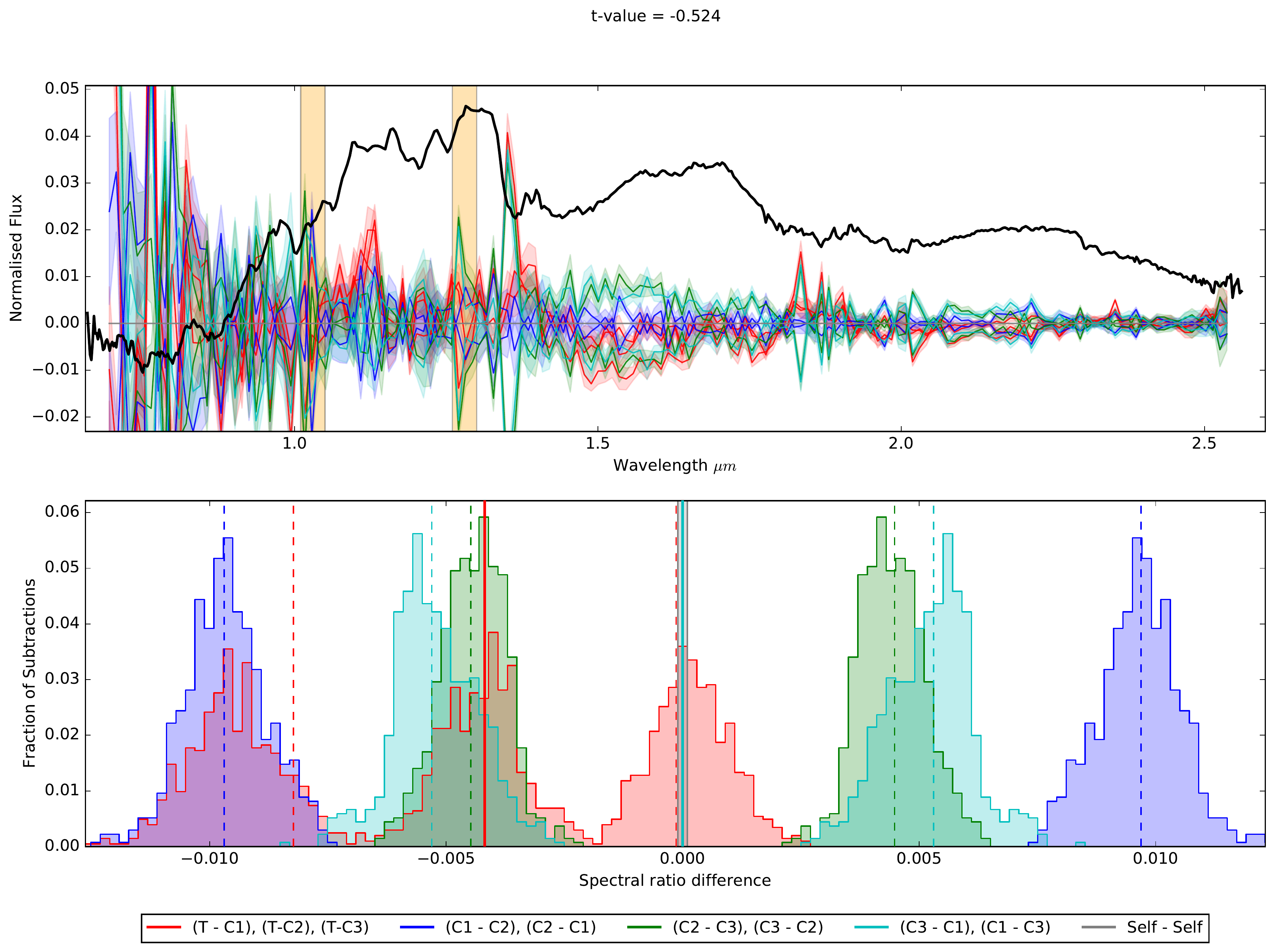}
        \\ (b) WISE J140145.91+310640.6, part of group 228.
        \end{center}
\end{minipage}
\caption{Layout identical to Fig. \protect \ref{figure:specdiff_for_simulation}. (a) An example of a target minus control subtraction, one of our M+UCD candidates for which a spectroscopic signature of a UCD was found (t-value $ = 7.8$) where one control star was rejected. (b) An example of a non-detection (t-value $ \sim 0.0$) where no control stars were rejected. Plotted in black is a comparison L0 residual (M4 is LP 508-14, \protect\citealt{Burgasser2004}, and the L0 is  2MASP J0345432+254023, \protect\citealt{BurgasserMcElwain2006}). \label{figure:specdiff_results}}
\end{figure*}

We first consider the four targets whose t-values are greater than 1.75, and look more closely for potential signatures of unresolved UCD companions. The most interesting candidate is WISE J100202.50+074136.3, which was analysed as part of group 800 and yielded a t-value of 7.28. The spectral residuals and spectral ratio difference histograms for this group are shown in Fig.\ref{figure:specdiff_results}a. The residuals show a distinct maximum across the 1--1.3 $\mum$ range (i.e. the region covered by our selected spectral ratio difference bands), and this maximum appears to be significantly greater than the level of scatter we might expect in the absence of a near-infrared excess. In the 1.3--1.8 $\mum$ range an excess signal is less clear, but may be present at a lower level. We also note a dip in the residuals at either end of this wavelength range, which is consistent with early L morphology. At longer wavelengths, any excess signal is lower still, but may be present when compared to the null case residuals. Overall the residuals are reasonably consistent with an early L dwarf across the full spectral range, and the significance of the 1--1.3 $\mum$ signal is born out in the spectral ratio difference histogram that shows the M+UCD candidate minus control values are well separated from the population of control star minus control star values. In contrast, Fig. \ref{figure:specdiff_results}b shows similar plots for a clear non-detection in our sample (WISE J140145.91+310640.6, part of group 228, t-value=--0.54). The mass of an $\sim$L0 companion to WISE J1002+0741 would depend on the age of the system. The latest BT-Settl models \citep{Baraffe2015} suggest that for ages from $\sim$0.6--2 Gyr the object would have a mass close to the hydrogen burning mass limit, and for older or younger ages it would be stellar or sub-stellar respectively. Age constraints are not currently available for WISE J1002+0741 however.

\clearpage
\newpage

\begin{table}
\caption{Rejection table for our colour-similar control stars. Those control stars in bold were rejected as having either $\chi^{2}_{\rm Red}$$>$5 (where the chi-squared fit is a comparison of the optical, $<1 \mum$, part of the target and control spectra) or \JKcut  $>2\sigma$ (a comparison between the similarity of the target and control using a spectral ratio with bands $1.01--1.05 \mum$ and $2.10--2.30 \mum$). Note Group 7 only had two control stars observed due to time constraints. \label{table-rejection_table}}
\begin{center}
\begin{tabular}{lcccccc}

\hline

Group & \multicolumn{2}{c}{Control 1} & \multicolumn{2}{c}{Control 2} & \multicolumn{2}{c}{Control 3} \\
number & $\chi^{2}_{\rm Red}$ & YK \% & $\chi^{2}_{\rm Red}$ & YK \% & $\chi^{2}_{\rm Red}$ & YK \% \\
\hline
7 & 0.55 & 6.3 & 0.66 & 4.6 &  &  \\
68 & 0.59 & 10.2 & 0.98 & 1.7 & 0.69 & 2.5 \\
92 & 1.87 & 9.3 & 0.67 & 3.6 & 1.17 & 0.9 \\
109 & 3.6 & 5.2 & 2.47 & 7.0 & {\bf 10.45} & {\bf 16.8} \\
124 & {\bf 1.68} & {\bf 17.6} & 1.04 & 10.0 & {\bf 0.55} & {\bf 11.4} \\
159 & 0.48 & 0.8 & {\bf 1.1} & {\bf 15.1} & 0.65 & 6.1 \\
228 & 0.5 & 3.7 & 0.43 & 3.7 & 1.04 & 2.0 \\
232 & 0.46 & 10.6 & 0.89 & 5.7 & {\bf 26.79} & {\bf 32.4} \\
282 & 0.53 & 7.7 & 0.5 & 5.0 & {\bf 0.62} & {\bf 11.9} \\
340 & {\bf 4.3} & {\bf 13.1} & 1.2 & 3.4 & 0.77 & 6.8 \\
360 & 0.53 & 5.4 & 0.45 & 2.1 & 0.47 & 1.4 \\
466 & 0.69 & 3.6 & 0.63 & 0.7 & 0.67 & 1.6 \\
476 & {\bf 2.23} & {\bf 15.5} & {\bf 1.19} & {\bf 12.1} & 0.8 & 6.3 \\
550 & 2.02 & 8.7 & 0.81 & 4.3 & 1.17 & 2.7 \\
697 & 0.92 & 1.9 & 1.02 & 5.0 & {\bf 24.71} & {\bf 26.0} \\
751 & 1.23 & 4.8 & 0.69 & 2.0 & 0.69 & 0.6 \\
757 & 0.98 & 9.7 & 0.73 & 6.0 & 1.36 & 6.3 \\
761 & 0.49 & 3.6 & 0.98 & 3.7 & 2.12 & 5.2 \\
800 & {\bf 2.5} & {\bf 12.0} & 2.32 & 2.6 & 1.02 & 4.2 \\
862 & 1.11 & 0.1 & 0.88 & 4.2 & 1.49 & 1.0 \\
890 & 1.3 & 1.3 & 0.8 & 9.7 & 0.93 & 3.8 \\
918 & 0.61 & 2.5 & 1.09 & 1.8 & 0.87 & 1.6 \\
927 & {\bf 0.61} & {\bf 11.8} & {\bf 1.25} & {\bf 13.1} & {\bf 2.1} & {\bf 13.3} \\
942 & 0.65 & 8.6 & 0.75 & 2.5 & {\bf 1.2} & {\bf 13.9} \\
946 & 0.66 & 2.9 & 0.53 & 1.5 & 1.01 & 0.9 \\
956 & 0.56 & 0.7 & 0.62 & 0.4 & 0.75 & 5.6 \\
961 & 0.35 & 2.1 & 0.4 & 0.0 & 0.45 & 2.9 \\
985 & 0.39 & 2.1 & 0.53 & 5.4 & 0.59 & 2.4 \\
\hline
\end{tabular}
\end{center}
\end{table}

\begin{table}
\caption{The resulting t-values for all observations. $T-C$ is target minus control subtractions, $M-C$ is the subtractions where a control star (to which a UCD has been added) is subtracted from the other control stars, and $CX-C$ is the subtractions of one control star (chosen as the target) minus the other control stars. In bold are those that we indicate as possible detections (t-value > 1.75). \label{ch4_table_tvalues_all}}
\begin{center}
\begin{tabular}{cccccc}
\hline
Group & $T--C$ & $M--C$ & $C1--C$ & $C2--C$ & $C3--C$ \\
\hline
7 & 0.99 & {\bf 2.73} & -- & -- & -- \\
68 & --0.39 & {\bf 2.08} & 0.01 & 1.25 & --1.33 \\
92 & 0.12 & 0.37 & --4.13 & 0.64 & 0.29 \\
109 & {\bf 6.44} & {\bf 6.75} & -- & -- & -- \\
159 & {\bf 6.61} & {\bf 6.36} & -- & -- & -- \\
228 & --0.54 & 1.45 & 1.51 & --1.3 & --0.03 \\
232 & --5.83 & {\bf 2.94} & -- & -- & -- \\
282 & 1.59 & 1.24 & -- & -- & -- \\
340 & 1.63 & 1.38 & -- & -- & -- \\
360 & --2.1 & {\bf 2.74} & --0.12 & 1.55 & --1.01 \\
466 & --0.05 & 1.28 & --2.86 & 0.22 & 0.71 \\
550 & 0.49 & 1.56 & --0.53 & --0.36 & {\bf 8.03} \\
697 & {\bf 2.09} & {\bf 1.76} & -- & -- & -- \\
751 & 0.44 & 1.05 & --5.99 & 0.4 & 0.42 \\
757 & --1.1 & 1.15 & 1.32 & 0.02 & --1.46 \\
761 & --0.47 & 0.73 & 0.81 & --2.39 & 0.18 \\
800 & {\bf 7.28} & {\bf 3.43} & -- & -- & -- \\
862 & 0.69 & 0.78 & --1.17 & 1.46 & --0.04 \\
890 & --2.3 & {\bf 5.12} & --0.05 & --0.87 & 1.31 \\
918 & --0.19 & 0.42 & --8.71 & 0.56 & 0.36 \\
942 & 0.16 & 0.79 & -- & -- & -- \\
946 & 0.36 & {\bf 2.25} & 0.85 & --1.86 & 0.27 \\
956 & 0.12 & --0.52 & 0.23 & 0.69 & --3.63 \\
961 & --0.5 & 0.51 & --6.72 & 0.36 & 0.56 \\
985 & --0.57 & 0.77 & 0.31 & --3.61 & 0.63 \\
\hline
\end{tabular}
\end{center}
\end{table}

\begin{figure*}
\begin{minipage}{.85\textwidth}
        \begin{center}
        \includegraphics[trim={0 0 0 2cm}, clip, width=\textwidth]{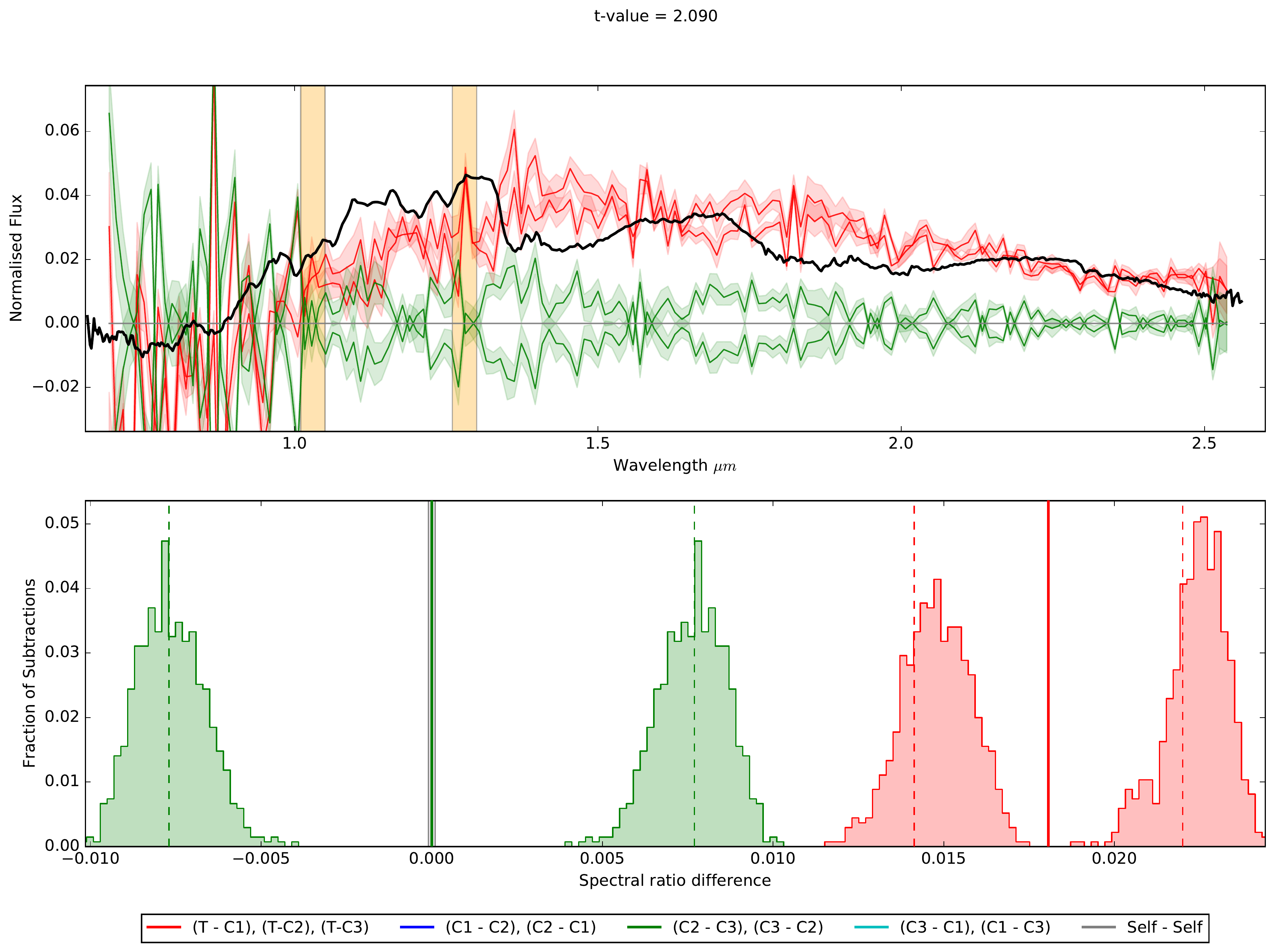}
        \\ (a) WISE J104507.41+181311.0, part of group 697.
        \end{center}
\end{minipage}
\begin{minipage}{.85\textwidth}
        \begin{center}
        \includegraphics[trim={0 0 0 2cm}, clip, width=\textwidth]{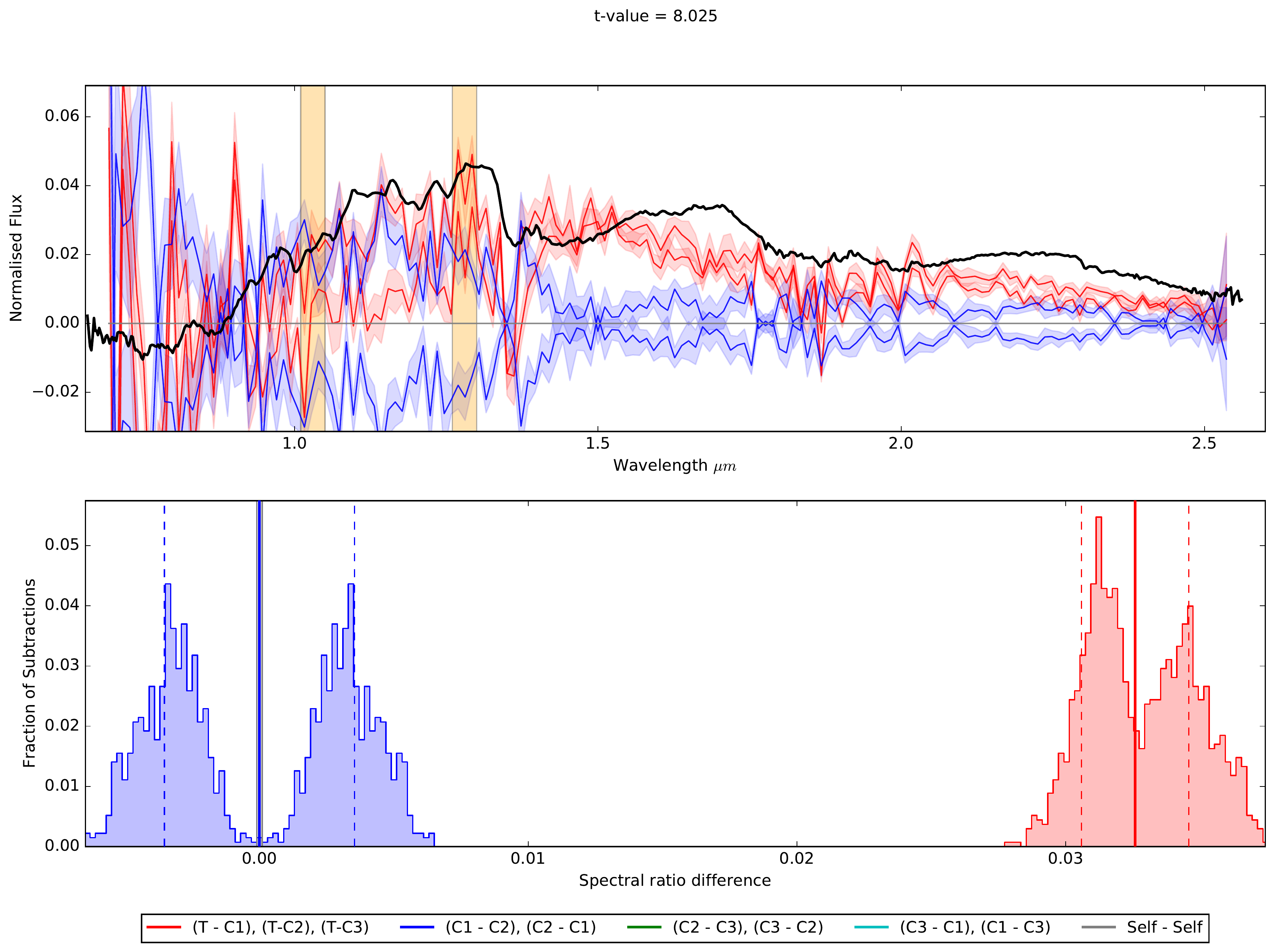}
        \\ (b) WISE J115522.06+002657.3, part of group 550.
        \end{center}
\end{minipage}
\caption{Layout identical to Fig. \protect\ref{figure:specdiff_for_simulation}. (a) An example of target minus control subtraction, showing significant red residuals but of a somewhat different morphology to WISE J100202.50+074136.3. (b) An example control minus control subtraction, this residual signal was the only one from our our type 2 analysis to yield a t-value$>$1.75. Plotted in black is a comparison L0 residual (M4 is LP 508-14, \protect\citealt{Burgasser2004}, and the L0 is  2MASP J0345432+254023, \protect\citealt{BurgasserMcElwain2006}). \label{figure:specdiff_results2}}
\end{figure*}

\begin{figure}
        \begin{center}
        \includegraphics[trim={17.5cm 0 0 17.5cm}, clip, width=.5\textwidth]{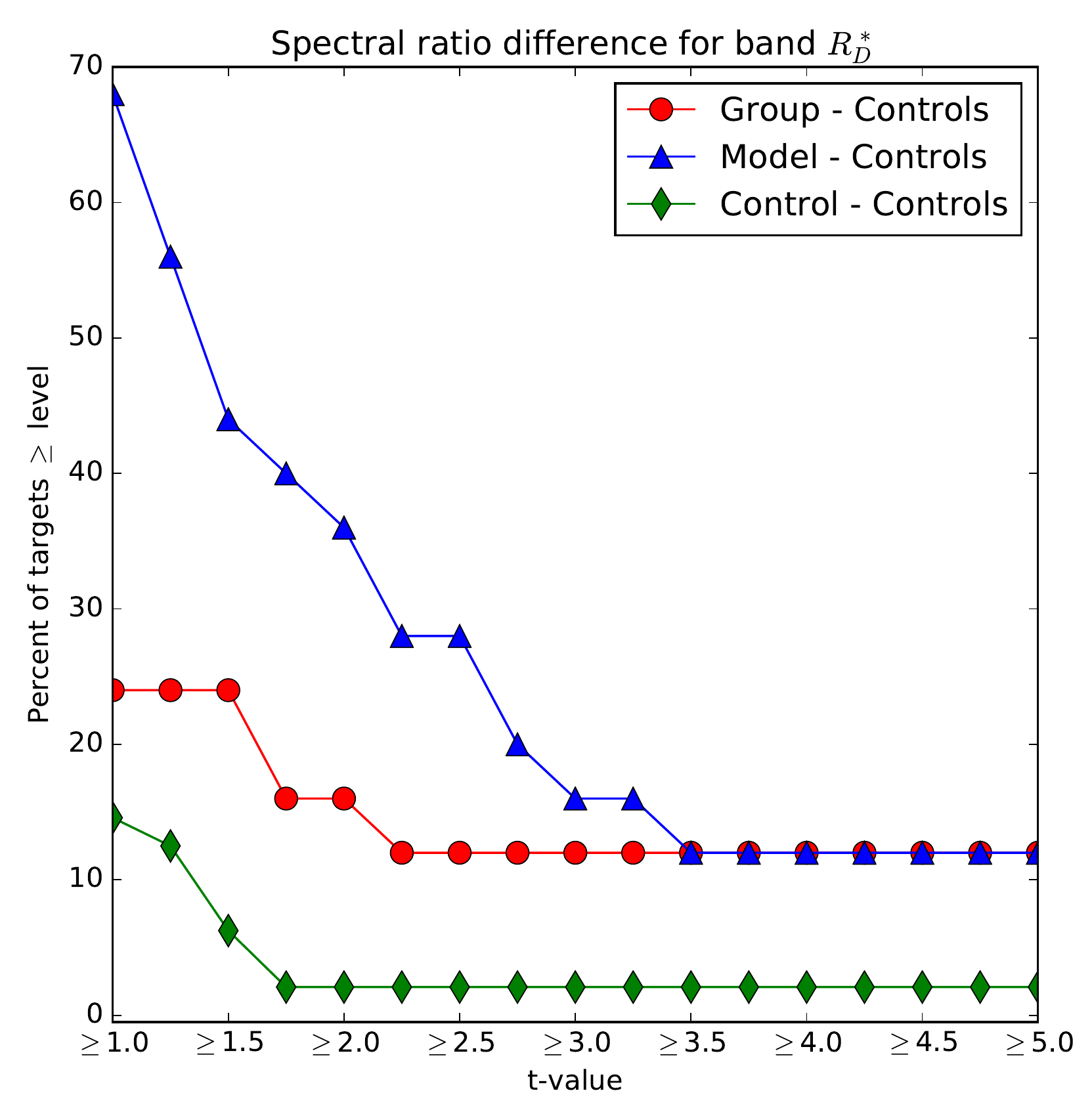}
        \end{center}
        \caption{Percentage of groups yielding a detected unresolved UCD signature as a function of t-value threshold. The three types of analysis (M+UCD candidate minus control star, control star minus control star and model candidate minus control star (see Section \ref{section:results}) are shown as red, green and blue symbols respectively. \label{figure:likelihood_compare}}
\end{figure}

Our other three possible detections come in two types. One (WISE J143046.74+272058.2, part of group 159) shows no apparent red residuals despite a significant t-value. Examination of the residual trace in the spectral ratio bands suggests this that t-value has been spuriously inflated by noise fluctuations and/or telluric absorption. The other two (WISE J104507.41+181311.0, part of group 697, and WISE J150642.41+324609.9, part of group 109) show significant red residuals, but of a somewhat different morphology to WISE J100202.50+074136.3. Fig. \ref{figure:specdiff_results2}a shows the red residuals for the first of these candidates as an example, and these residuals peak around a wavelength of 1.6 $\mum$, falling off on either side. In Fig.\ref{figure:specdiff_results2}b we provide a comparison spectral residual from our type 2 (control star minus control star) analysis. This residual signal was the only one from our type 2 analysis to yield a t-value$>$1.75, and is thus our best available example of a false positive. We note significant similarity between the example candidate (Fig.\ref{figure:specdiff_results2}a) and the false positive (Fig.\ref{figure:specdiff_results2}b), both of which have excess morphology that peaks around 1.6 $\mum$. It thus seems likely WISE J104507.41+181311.0 and WISE J150642.41+324609.9 do not have unresolved UCD companions, although the excess flux (seen in their residual spectra) may well be genuine but with a different origin.

In order to broadly assess ultracool companionship and contamination in our full target sample we now discuss the outcome of the three different types of analysis discussed in Section \ref{section:results}. Fig.\ref{figure:likelihood_compare} shows the percentages of M dwarfs with t-values greater than a range of different levels. The red, green and blue symbols/lines are for type 1, 2 and 3 analysis (\ie t-values calculated from spectral ratio differences between M+UCD candidate and control star pairs, pairs of control stars, and model candidate and control star pairs). It is clear from the green symbols our method only identifies false positive signals around a small fraction of our control star pairings (just one with a t-value $>$1.75). However, this signal can have quite a high significance, and may well represent a genuine excess albeit not due to an unresolved UCD companion. The blue symbols show if every M dwarf had an unresolved UCD companion then they would manifest as a range of t-values, more numerous for lower values, and with an expected yield of $\sim$40 per cent with a t-value $>$1.75. The red symbols lie about mid-way between the blue and green for t-values $<$3.5, but lie close to the blue symbols (and well above the green symbols) for higher t-values. We loosely interpret this as being consistent with an intermediate percentage of unresolved UCD companions, some fraction of which should have t-value $>$1.75, combined with several times the number of false positives present amongst the control stars. This is roughly consistent with the results suggested by our spectral residual inspection (\ie we have one good unresolved UCD candidate and
three probable false positives).

We have considered a variety of possible effects that might lead to false positive signatures amongst our candidates, and estimated their likelihood and/or expected levels of occurrence.

Intrinsic properties of the target M dwarf systems are an important consideration. M dwarf variability could lead to an apparent mid-infrared (MIR) photometric excess (leading to an M dwarf being selected in our \citetalias{Cook2016} analysis), and if such variability is due to spots and rotation then this might also affect their spectra \eg high level spot coverage could lead to a spectrum consisting of warmer (photospheric) and cooler (from spots) components. However, our \citetalias{Cook2016} analysis established the M+UCD candidates are not significantly variable on time-scales of hours to days. In any event photosphere/spot temperature differentials for M dwarf spots are of order $\sim$150 K \citep{Vida2016} and thus unlikely to mimic the signature of an unresolved UCD companion. An unresolved M (non-UCD) companion could lead to low level differences between a candidate and a single M dwarf control star, however the residual spectra would not resemble a UCD and the resulting MIR excess would be negligible. The presence of dusty discs around M dwarfs could cause a MIR excess, however an NIR/MIR multicolour analysis of our candidate sample (Appendix \ref{section:excess_contamination:discs}) indicates there are no obvious discs present that could account for the unresolved companion signatures.

Chance alignments, where background/foreground objects are blended with the candidates, could produce MIR excess and affect the NIR spectral shape. However, we have estimated likely levels of such alignments (Appendices \ref{section:excess_contamination:red_objects}--\ref{ch4_section_chance_align_galaxies}) and expect no more than 0.1 per cent due to galactic sources (cooler M dwarfs, M giants and brown dwarfs), and a worst case scenario of 8--9.5 per cent due to red galaxies. Indeed, a random-offset-analysis suggests an $\sim$3.5 per cent level of chance alignments with sufficient excess to be mistaken for a UCD (Appendix \ref{ch4_section_chance_alignment_from_random_offsets}).  We also note we have visually inspected candidates in WISE, 2MASS, SDSS and where possible in UKIDSS and DSS2 (Appendix \ref{ch4_section_visual_inspection}) to rule out source blending as much as possible. While low level of blending may remain (causing the MIR excess of some candidates), any resulting NIR excess would not be expected to resemble UCD morphology. 

We also modelled the effects in residual spectra using M dwarf pairings with slightly different properties (to see if false positives could be generated in the absence of a UCD companion). Using 75 PHOENIX synthetic spectra \citep{Husser2013} with combination of \Teff (2800, 2900, 300, 3100, 3200 K), log g (4.5, 5.0, 5.5) and $[$Fe$/$H$]$ (--1, --0.5, 0, 0.5, 1), we generated 5500 residual spectra that were subjected to our M+UCD candidate analysis procedures. Of these 884 (16 per cent) passed the spectral similarity criteria ($\chi^{2}_{\rm Red}$$>$5  or \JKcut  $>2\sigma$) and 417 (7.6 per cent) yielded t-values greater than 1.75. However, the relative \JWb excess between these pairings was in the range 0.002$\pm$0.026 (3$\sigma$), well below the level required to be selected as M+UCD candidates in the first place. By comparison WISE J100202.50+074136.3 has a relative \JWb excess of 0.093 mag and our full M+UCD candidate sample has a minimum excess of $\sim$0.045 mag (see \citetalias{Cook2016}).

Overall this analysis suggests an expected NIR excess rate of 3.5--9.5 per cent, due mainly to blended background galaxies. This compares favourably with the two candidates we identify as probably false positives in our analysis (showing NIR excess in their residual spectra, but with non-UCD-like morphology). We have not found a good alternative explanation for WISE J100202.50+074136.3, and consider it to be a strong M+UCD candidate.

Using the results from our first follow-up sample we estimated the occurrence rate of unresolved UCD companions to M dwarfs in our full candidate sample, as well as the potential `catch' that would result from comprehensive follow-up. Based on our assessment our sample contains one unresolved UCD companion and three false positives, combined with an approximate recovery rate of $\sim$40 per cent (from the model candidate and control star results in Fig.\ref{figure:specdiff_results2}) we estimate an occurrence rate of $\frac{1}{24\times 0.4} \sim 0.1$. This is substantially higher than the expected occurrence rate for a randomly selected sample of M dwarfs, and although this is based on just one good candidate it suggests our candidate selection method (from paper 1) may be achieving its desired goal. If this occurrence rate is appropriate for our full excess sample of 1 082 candidates then we might expect up to $\sim$100 unresolved UCD companions within the sample, and the potential to recover $\sim$40 of these through an expanded spectroscopic follow-up campaign (c.f. Fig.\ref{figure:mass_vs_separation}, where only a handful of such M3-M5 companions are known).

    \section{Conclusions}
        \label{section:conclusions}

We have developed a spectroscopic method to identify the signatures of unresolved L dwarf companions to mid-M dwarfs, with targets and their associated optically colour-similar control stars coming from the photometric analysis of \citetalias{Cook2016}. As a first stage, our method makes use of spectral ratio differences between the spectra of candidates and their control stars, which mitigates against the scatter in M dwarf colours (and spectral morphology) that occurs across the full population for any particular M spectral type. As a second stage our method examines spectral difference residuals (between candidate and control star pairings), to visually reveal any near-infrared excess flux from ultracool companions. Testing showed our spectroscopic method is optimized by spectral ratio differences in the 1.21--1.35 $\mum$ and 0.96--1.10 $\mum$ bands, and the best near-infrared spectroscopy for this purpose should have a spectral resolution of $\sim$200 and a signal to noise of $\sim$125--200. 

We obtained a suitable data set for a pilot sample from \citetalias{Cook2016} during good conditions with SpeX on the IRTF. The identification of the strong signature for WISE J100202.50+074136.3 (\JWb excess of 0.093), was recognized with a t-value of 7.28 and showed early L-like morphology in the residual spectra, is encouraging and should be followed up. Adaptive optics should be capable of resolving a companion at separations $>$0.1 \arcseconds ($>$15 au at the M dwarf distance of 150 pc), and radial velocity variations\footnote{Calculated online at \url{http://astro.unl.edu/classaction/animations/extrasolarplanets/radialvelocitysimulator.html}} (of at least $\sim$3 km s$^{-1}$) would be expected for separations out to $\sim$1 au (with periods ranging up to $\sim$2 yr). Also, {\it Gaia} (during its 5 yr mission) may detect an astrometric wobble of several milli-\arcseconds amplitude for separations out to $\sim$3 au. Thus direct/indirect detection of this candidate companion may be eminently possible with current facilities. And with full follow-up of our candidate sample from \citetalias{Cook2016}, we might expect to confirm up to $\sim$40 such companions in the future, adding extensively to the known desert population of M3-M5 dwarfs.

\section*{Acknowledgements}

\small 
NJC acknowledges support from the UK's Science and Technology Facilities Council (grant number ST/K502029/1), and has benefited from IPERCOOL, grant number 247593 within the Marie Curie 7th European Community Framework Programme. FM acknowledges support from the UK's Science and Technology Facilities Council (grant number ST/M001008/1). Support for RGK is provided by the Ministry for the Economy, Development, and Tourisms Programa Inicativa Cientifica Milenio through grant IC 12009, awarded to  The  Millennium  Institute  of  Astrophysics (MAS) and acknowledgement to CONICYT REDES No. 140042 project. R.G.K. is supported by Fondecyt Regular No. 1130140. Visiting Astronomer at the Infrared Telescope Facility, which is operated by the University of Hawaii under contract NNH14CK55B with the National Aeronautics and Space Administration. We make use of data products from WISE \citep{Wright2010}, which is a joint project of the UCLA, and the JPL$/$CIT, funded by NASA and 2MASS \citep{Skrutskie2006}, which is a joint project of the University of Massachusetts and the Infrared Processing and Analysis Center$/$CIT, funded by NASA and the NSF. We also make substantial use of SDSS DR10, funding for SDSS-III has been provided by the Alfred P. Sloan Foundation, the Participating Institutions, the NSF, and the USDOESC. This research has made use of the NASA$/$IPAC Infrared Science Archive, which is operated by JPL, CIT, under contract with NASA, and the VizieR data base catalogue access tool and SIMBAD data base\citet{Wenger2000}, operated at CDS, Strasbourg, France. This work is based in part on services provided by the GAVO Data Center and the data products from the PPMXL data base of \citet{Roeser2010}. This publication has made use of LAMOST DR1 and DR2 spectra. Guoshoujing Telescope (LAMOST) is a National Major Scientific Project built by CAS. Funding for the project has been provided by the National Development and Reform Commission. LAMOST is operated and managed by the NAO, CAS. This research has benefited from the SpeX Prism Spectral Libraries, maintained by Adam Burgasser. This research made extensive use of: {\sc Astropy} \citep{Astropy2013}; {\sc matplotlib} \citep{Chabrier2007}, {\sc scipy} \citep{jones2001}; {\sc Topcat} \citep{Topcat}; {\sc Stilts} \citep{Stilts} {\sc ipython} \citep{Perez2007} and NASA's Astrophysics Data System. We thank the anonymous reviewer for their careful reading of our manuscript and for the insightful comments and suggestions made.

\bibliographystyle{mnras}

\begin{appendix}

\section{Contamination in the excess sample and candidate M+UCDs}
    \label{section:excess_contamination}

\subsection{Contamination from discs}
\label{section:excess_contamination:discs}

Circumstellar discs around M dwarfs can be approximated quite well by a blackbody of temperature, \Teff and of extent $R$. These discs are heated by the central star and as such cannot exceed the stellar temperature unless some other process is involved. One way discs are found is to look for MIR excess (for M dwarfs \eg  \citealt{Esplin2014}, \citealt{Theissen2014} and \citealt{Luhman2012b}). 

\citet{Esplin2014} investigate the excess signal via MIR continuum emission from warm circumstellar dust. They define a boundary \citep[see Fig. 2 from][]{Esplin2014} above which stars have an excess signal due to this warm dust. \citet{Theissen2014} compliment this with a polynomial fit \citep[see Table 1 ][]{Theissen2014} to the M dwarf main sequence with \WaWc and \WbWc as functions of \RZ from SDSS. Our M dwarfs lie well below the region in which M dwarfs  are known to have circumstellar discs lie (Fig. 3 from \citealt{Theissen2014}), and mostly lie out of the region defined by \citet{Esplin2014}. \citet{Luhman2012b} also present a boundary in \KWb colour-spectral type space, however their boundaries lies significantly above our distribution and the boundary of \citet{Esplin2014}, thus having no overlap with our M+UCD candidates. 

    \begin{figure*}
        \begin{center}
        \includegraphics[width=.9\textwidth]{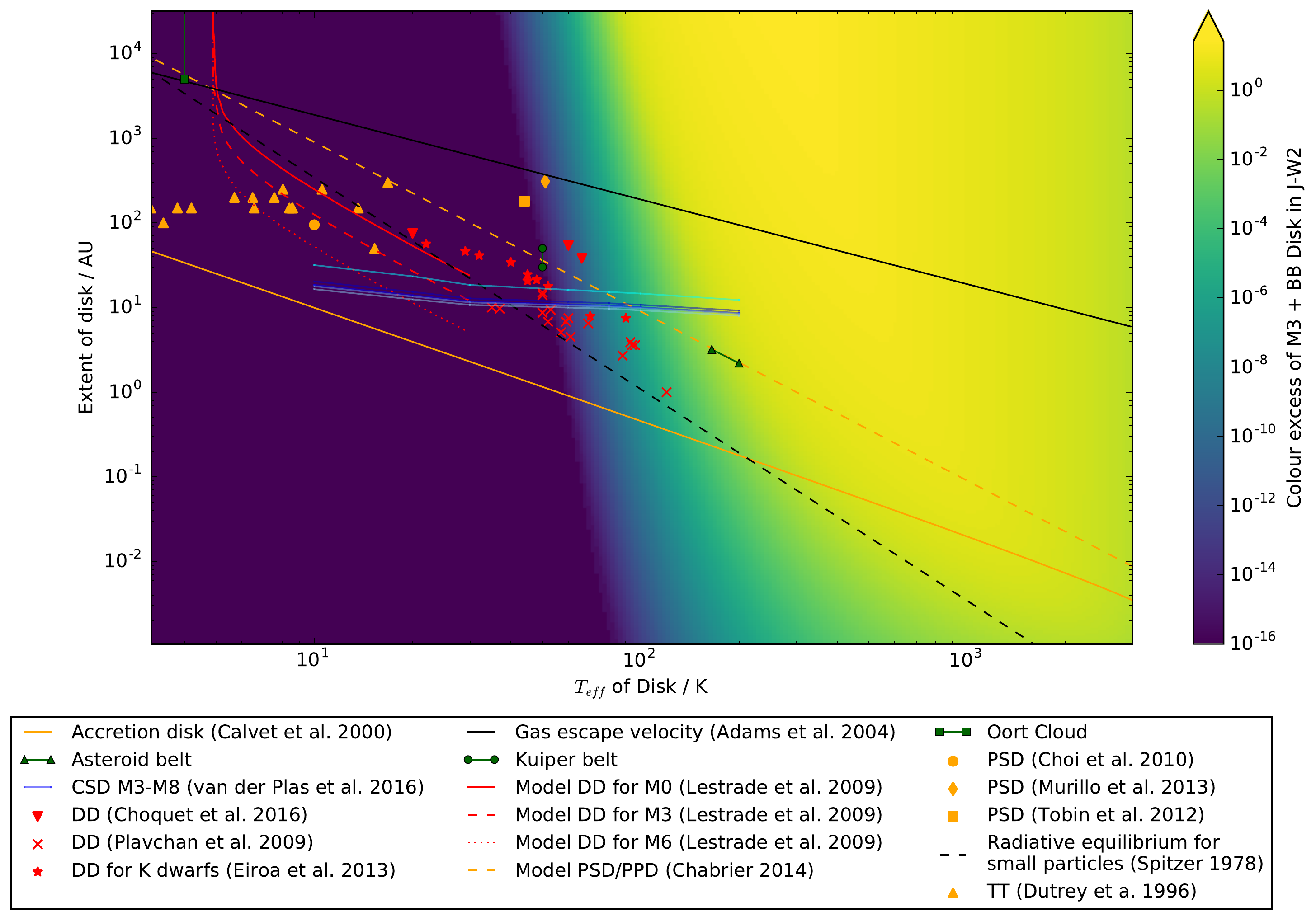}
        \end{center}
        \caption[Simulations of the colour excess from an M dwarf with an added blackbody of temperature, \Teff and surface area $=\pi(\text{extent})^2$ for colour excess in \JWb.]{Simulations of the colour excess from an M dwarf with an added blackbody of temperature, \Teff and surface area $=\pi(\text{extent})^2$ for colour excess in \JWb. Over plotted are literature examples of circumstellar discs, CSD, debris discs, DD, protostellar discs, PSD and T-Tauri stars, TT (\citealt{Dutrey1996}, \citealt{Plavchan2009}, \citealt{Choi2010}, \citealt{Tobin2012}, \citealt{Eiroa2013}, \citealt{Murillo2013}, \citealt{Choquet2016}) and models (\citealt{Spitzer1978}, \citealt{Calvet2000}, \citealt{Adams2004}, \citealt{Lestrade2009}, \citealt{Chabrier2014}, \citealt{vanderplas2016}). Note the Solar system objects (Asteroid belt, Kuiper belt and Oort Cloud) are plotted as comparisons to warmer stars and as such M dwarf levels of excess would be much lower. \label{ch4_figure_discexcess_jw2}}
    \end{figure*}

    To investigate the effects of circumstellar reddening further, we investigated ways in which warm discs (of various size) might give excess values that could contaminate our selection of M+UCD candidates.

    \begin{equation}
    \label{ch4_equation_bb}
    \begin{split}
    & B_{\lambda} = \frac{2hc^2}{\lambda^5} \frac{1}{\text{exp}(hc/\lambda k_B \Teff)-1} \\
    & F_{\lambda} = \pi B_{\lambda} \\
    & F_{\lambda}(d) = F_{\lambda} \frac{\Sigma_{\text{disc}}}{\Sigma_{\text{sphere}}}
    \end{split}
    \end{equation}

    \noindent where $B_{\lambda}$ is the spectral radiance, $F_{\lambda}$ is the flux, $F_{\lambda}(d)$ is the flux as observed from a distance $d$, $\Sigma_{\rm disc} = \pi(R_{\rm outer}^{2} - \pi R_{\rm inner}^{2})$ is the surface area of the disc with inner radius $R_{\rm inner}$ and outer radius $R_{\rm outer}$ and $\Sigma_{\text{sphere}} = 4\pi d^{2}$ is the surface area of a sphere at radius $d$, the distance of observation (taken to be 10 pc).

    We made a grid of 250 values of $0.5 < \text{log}(\frac{\Teff}{K}) < 3.6$ and 250 values of $-3.5 < \text{log}(\frac{\text{extent of disc}}{au}) < 4.5$ (where $\text{extent of disc} \approx R_{\rm outer}$ as we set $R_{\rm inner} = R_{*}$). From these 62 500 \Teff and extent of disc combinations we added a blackbody as described in \refequ{ch4_equation_bb} to the M dwarf BT-Settl \citep[CIFIST2011\_2015][]{Baraffe2015}\footnote{Accessed online at \url{https://phoenix.ens-lyon.fr/Grids/BT-Settl/}} model (smoothed to 5 000 bins for faster computation between 0 and 30 $\mum$, and normalized to 10 pc). For each point in the grid the colour excess ($\text{Colour}_{\text{M}+\rm disc} - \text{Colour}_{\text{M dwarf}}$) was calculated for \JH, \HWa, \HWb, \JWa, \JWb and \WbWc where colour is calculated in \refequ{ch4_equation_colour_from_spectra}.

    \begin{equation}
    \begin{split}
    \label{ch4_equation_colour_from_spectra}
    \text{Colour}(1, 2) & = M_1 - M_2 \\
    & = -2.5 \text{log}_{10}\left( \frac{\int{I_{\lambda}\tau_{1}(\lambda)\text{d}\lambda}}{\int{I_{0,1}\tau_{1}(\lambda)\text{d}\lambda}} \frac{\int{I_{0,2}\tau_{2}(\lambda)\text{d}\lambda}}{\int{I_{\lambda}\tau_{2}(\lambda)\text{d}\lambda}} \right)
    \end{split}
    \end{equation}

    \noindent where $I_{\lambda}$ is the flux from the spectrum, $\tau_{1}(\lambda)$ is the transmission profile of band 1 and $I_{0,1}$ is the zero-point flux of band 1\footnote{2MASS bands from \url{http://www.ipac.caltech.edu/2mass/releases/allsky/doc/sec6_4a.html} and WISE bands from \url{http://wise2.ipac.caltech.edu/docs/release/allsky/expsup/sec4_4h.html}}.

    To these grids, we added data from the literature for known circumstellar discs, CSD, debris discs, DD, around low-mass stars (from \citealt{Plavchan2009}, \citealt{Eiroa2013}, \citealt{Choquet2016}) protostellar discs, PSD, around low-mass stars (from \citealt{Choi2010}, \citealt{Tobin2012}, \citealt{Murillo2013}), low-mass T-Tauri stars, TT (from \citealt{Dutrey1996}) and from the Solar system (Asteroid belt, Kuiper belt and Oort cloud). We also added some models of debris discs from \citet{Lestrade2009} for M0, M3 and M6 dwarfs; a model of accretion discs from \citet{Calvet2000}, models of circumstellar discs for M3--M8 dwarfs from \citet{vanderplas2016}, a model of gas escape velocity from \citet{Adams2004}, a model of PSD and proto-planetary discs, PPD from \citet{Chabrier2014} and a model of radiative equilibrium for small particles by \citet{beckwith1990} and \citet{Spitzer1978}.

    We used these to show how much colour excess at given \Teff and extent would add to an M dwarf assuming they were blackbodies. Fig. \ref{ch4_figure_discexcess_jw2} shows these grid points plotted for an M4 dwarf with the comparison to the literature.

    \begin{figure}
        \begin{center}
        \includegraphics[width=.45\textwidth]{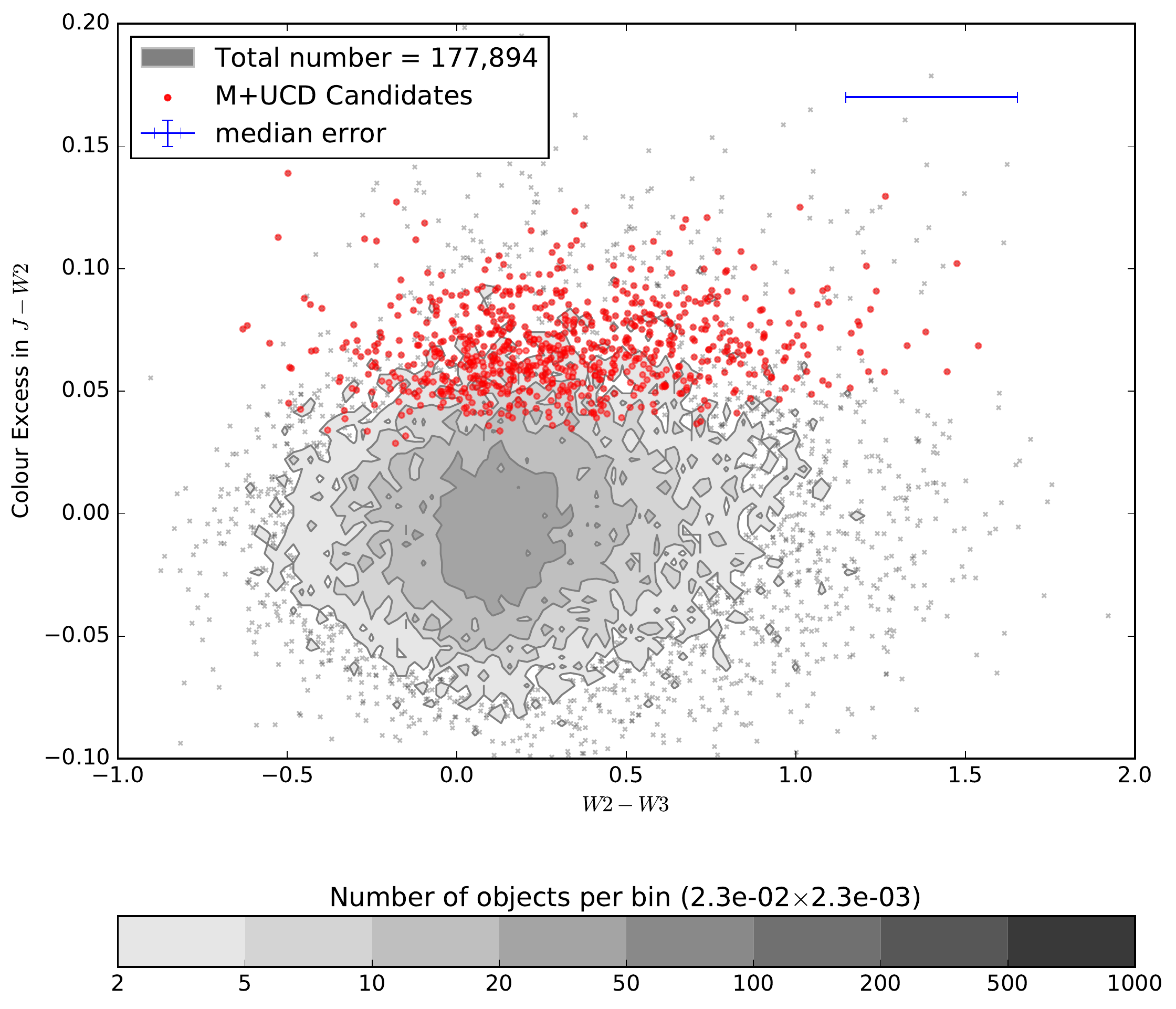}
        \end{center}
        \caption{Colour excess in \JWb against \WbWc for all our M dwarfs in the full M dwarf candidate catalogue and for our M+UCD candidates that have a $W3$ detection. This plot shows the distribution of our M+UCD candidates is consistent with our over all distribution of M dwarfs in \WbWc and thus there are no obvious candidates which have extremely large \WbWc colour (\ie there are no signatures of a circumstellar discs in our M+UCD candidates). \label{ch4_figure_w2w3_vs_excess_in_jw2}}
    \end{figure}

    As a red \WbWc colour is a clear signature of a disc we also plotted \WbWc against colour excess in \JWb for all our M dwarfs in the full M dwarf candidate catalogue with a $W3$ detection (non-upper limit) and for our M+UCD candidates that have a $W3$ detection (see Fig.\ref{ch4_figure_w2w3_vs_excess_in_jw2}). Fig. \ref{ch4_figure_w2w3_vs_excess_in_jw2} shows there are no major outliers and therefore no obvious discs present in our M+UCD candidates.

    In addition to this analysis, discs around late-K and M dwarfs seem to be rare and only present around young M dwarfs \citep[see ][ and references therein]{Deacon2013}. Even if these rare discs exist from work presented above it is clear only exceptionally warm or large discs would give the colour excess required to be mistaken for one of our M+UCD candidates.

   \subsection{Contamination from chance aligned red objects}
    \label{section:excess_contamination:red_objects}

   Foreground or background objects that appear redder than our M dwarfs and are randomly aligned within the WISE PSF will cause an M dwarf to look redder. We explore the various red objects, foreground and background objects that can redden our M dwarfs; we look at foreground and background M dwarf and brown dwarfs, background giants and galaxies. Although line-of-sight dust (local reddening) could also redden our M dwarfs, it is not clear how this can be easily modelled so we do not attempt this.

    To calculate how many reddened objects are expected to contaminate our M dwarfs we defined a spherical cone \citep{WeissteinSphericalCone} with volume, Vol. Any objects inside this spherical cone (centred around our M dwarf) appear blended due to the size of the M dwarfs PSF. This cone can also be used to calculate the density of objects in a certain area of sky, given a magnitude limit (and hence maximum distance; see \refequ{ch4_equation_vol_sphericalcone}.

    \begin{equation}
    \begin{split}
    \label{ch4_equation_vol_sphericalcone}
    & \text{Vol} = \frac{2}{3} \pi R^2 h = \frac{2}{3} \pi R^3 (1 - \text{cos}(\theta)) \\
    & \rho = \frac{N}{\text{Vol}} = \frac{N}{(\frac{2}{3} \pi R^3 (1 - \text{cos}(\theta))} \\
    \end{split}
    \end{equation}

    \noindent where Vol is the volume of the spherical cone , $R$ is the distance from the observer to the limit of visibility, $N$ is the number of objects in the volume and $\theta$ is the angular size of the cone.

    As well as being present within the PSF of WISE, objects need to contribute a sufficient amount of flux to give an excess in \JWb similar to adding a UCD to our M dwarfs. We define sufficient \JWb excess to mean 5 per cent colour excess \citepalias[see figures 6 and 7 from][]{Cook2016}. Using the definition of the colour excess and adding our target M dwarf and red contaminating object in flux space  leads to \refequ{ch4_equation_limiting_mag}, which is the limiting magnitude given a specific M dwarf and a specific \JWb of the contaminating object which can lead to a \JWb colour excess of, $E\JWb$.

    \begin{equation}
    \label{ch4_equation_limiting_mag}
    \begin{split}
    & E(J-W2) = (m_{J(M+B)} - m_{W2(M+B)}) - (m_{J(M)} - m_{W2(M)}) \\ 
    & m_{J(M+B)} = -2.5log_{10}(10^{-0.4m_{J(M)}} + 10^{-0.4m_{J(B)}}) \\ 
    & m_{W2(B)} = -2.5log_{10}\left( \frac{10^{-0.4(m_{J(M)}-E(J-W2) )} -10^{-0.4m_{W2(M)}}}{1-10^{-0.4((J-W2)_{B} - E(J-W2)  - (J-W2)_{M})}} \right)
    \end{split}
    \end{equation}

    \noindent where $m_{J(M+B)}$ and $m_{W2(M+B)}$ are the $J$ and $W2$ magnitudes of the combined M dwarf and red contaminating object, $m_{J(M)}$ and $m_{W2(M)}$ are the $J$ and $W2$ magnitudes of a specific M dwarf target, $\JWb_{B}$ is the colour of the red contaminating object and $m_{J(B)}$ is the limiting magnitude of the red contaminating needed to cause an Excess of $E(\JWb)$ ($\sim$0.05 for our sample).

    Using \refequ{ch4_equation_limiting_mag} we calculated the limiting magnitude a red contaminating object would need to have to sufficiently redden one of the M dwarfs. We chose not to apply extinction to the limiting magnitude due to the small amount of difference this would make (with an $A_V < 0.08$ and mean values of $\frac{A_{\lambda}}{A_V}$ of 0.179 for $J$ and 0.056 for $W2$, see Table B1 \citepalias{Cook2016}, the extinction is of order 0.01 in $J$ and 0.005 in $W2$).

    \subsection{Chance alignment of brown dwarfs and M dwarfs}
    \label{ch4_section_chance_align_bdm}

    For chance alignments of brown dwarfs and M dwarfs we took spatial densities from the literature (\citealt{PhanBao2003}, \citealt{Cruz2007}, \citealt{Reid2007}, \citealt{PhanBao2008}, \citealt{Burningham2013}, \citealt{Marocco2015}) and calculated the \JWb colour of the M dwarfs and brown dwarfs from our simulated photometry (see section 3.2 \citetalias{Cook2016}), these values are presented in Table \ref{ch4_table_bdm_spatial_densities_and_jw2_colour}.

    Using \JWb from Table \ref{ch4_table_bdm_spatial_densities_and_jw2_colour} we calculated the maximum distance brown dwarfs and M dwarfs could add sufficient flux to our target M dwarf (using \refequ{ch4_equation_limiting_mag}. This value was used if it was brighter than the limiting magnitude in 2MASS or WISE, otherwise the 1.25$\sigma$ 2MASS/WISE limit ($18.05$/$17.00$) was used instead\footnote{see Table 1 from \citetalias{Cook2016} where we add 1.5 to convert from 5$\sigma$ to 1.25$\sigma$ \label{footnote:convert_5_to_1.25}}. This number was calculated for each of the 36 898 M dwarfs in our excess sample. Then using \refequ{ch4_equation_vol_sphericalcone} we estimated the number of brown dwarfs or M dwarfs chance aligned with each of our target M dwarfs. Taking the sum of the number of objects for each of our target M dwarfs gave the total number of contaminating brown dwarfs and M dwarf expected in our M+UCD candidates. Our M+UCD candidates occupy an excess region between 0.05 and 0.15 in colour excess of \JWb. Therefore, we also subtract off those objects that have an excess greater than 0.15 to give a final number of objects. The results are shown in Table \ref{ch4_table_bdm_spatial_densities_and_jw2_colour} and we expect a total of no more than two of our 36 898 excess sample M dwarfs to be reddened due to a chance alignment with a foreground or background M dwarf or brown dwarf.

    \begin{table*}
    \caption{Results of the chance alignments of M dwarfs and brown dwarfs. Shown are the simulated M dwarf absolute photometry and \JWb colour (taken from section 3.2 of \citetalias{Cook2016}), and the spatial densities taken from the literature (a. \protect\citealt{Reid2007}, b. \protect\citealt{PhanBao2003}, c. \protect\citealt{PhanBao2008}, d. \protect\citealt{Cruz2007}, e. \protect\citealt{Marocco2015}, f. \protect\citealt{Burningham2013}). Note space densities from \protect\citet{Marocco2015} are stated for binary fractions of 26$\pm$13 and 14$\pm$10 respectively, and space densities from \protect\citet{Burningham2013} are stated from a minimum to maximum value and we take the worst case scenario in each case to calculate the contamination. Also calculated are the limiting magnitudes to give a 5 per cent excess in \JWb and thus the number of chance alignments of brown dwarfs and M dwarfs per target M dwarf and in total for our excess sample. Note that if the limiting magnitude was greater than the 1.25$\sigma$ detection limit of 2MASS or WISE, the 2MASS/WISE limit ($18.05$/$17.00$) was used instead$^{\ref{footnote:convert_5_to_1.25}}$. \label{ch4_table_bdm_spatial_densities_and_jw2_colour}}
    \begin{center}
\begin{center}
\begin{tabular}{cccccccccc}
\hline
spectral type & $M_J$ & $M_{W2}$ & \JWb & spatial density & $M_{J(\text{limit})}$ & $M_{W2(\text{limit})}$ & $\rho^{1}$ & T$^{2}$ & Ref. \\
 & mag & mag & mag & $\times10^{-3}$ pc$^{-3}$ & mag & mag & & & \\
\hline 
M2 -- M5   &   7.0 --  9.5  &    5.9 --  8.5  &  1.1 -- 1.3   &   76  &   12.39 -- 13.87   &   11.29 -- 12.57   &   1.20$\times10^{-4}$   &   1.59 &   a \\

M6 -- M8   &  10.3 -- 11.0  &    9.0 --  9.4  &  1.3 -- 1.6   &   4.62  &   13.87 -- 16.05   &   12.57 -- 14.45   &   4.41$\times10^{-6}$   &   0.18 &  b  \\

M8 -- L3.5 &  11.0 -- 12.9  &    9.4 -- 10.6  &  1.6 -- 2.3   &   3.28  &   16.05 -- 17.70   &   14.45 -- 15.39   &   3.13$\times10^{-6}$   &   0.12 &   c \\

L0 -- L3   &  11.6 -- 12.9  &    9.8 -- 10.9  &  1.8 -- 2.3   &   1.7$\pm$0.4  &  16.69 -- 17.70   &   14.89 -- 15.39   &   1.70$\times10^{-6}$   &   0.05 &   d \\

L4 -- L6.5 &  13.1 -- 14.1  &   10.7 -- 11.1  &  2.4 -- 3.0   &   0.85$\pm$0.55 and 1.00$\pm$0.64  &   17.85 -- 18.68   &   15.43 -- 15.68   &   6.30$\times10^{-7}$   &   0.02 &   e \\

L7 -- T0.5 &  13.3 -- 14.8  &   11.2 -- 11.8  &  3.1 -- 3.0   &   0.73$\pm$0.47 and 0.85$\pm$0.55  &   18.80 -- 18.70   &   15.70 -- 15.68   &   3.79$\times10^{-7}$   &   0.01 &   e \\

T1 -- T4   &  14.7 -- 14.8  &   11.9 -- 12.5  &  2.8 -- 2.3   &   0.74$\pm$0.48 and 0.88$\pm$0.56  &   18.42 -- 17.69   &   15.62 -- 15.39   &   1.33$\times10^{-7}$   &   0.00 &   e \\

T6 -- T6.5 &  14.7 -- 15.3  &   12.6 -- 13.0  &  2.1 -- 2.3   &   0.39$\pm$0.22 to 0.71$\pm$0.40  &   17.34 -- 17.69   &   15.24 -- 15.39   &   2.43$\times10^{-8}$   &   0.00 &   f \\

T7 -- T7.5 &  15.6 -- 16.0  &   13.1 -- 13.3  &  2.5 -- 2.7   &   0.56$\pm$0.32 to 1.02$\pm$0.64  &   18.01 -- 18.29   &    15.51 -- 15.59   &   2.51$\times10^{-8}$   &   0.00 &   f \\

T8 -- T8.5 &  16.6 -- 17.2  &   13.4 -- 13.6  &  3.2 -- 3.6   &   2.05$\pm$1.21 to 3.79$\pm$2.24  &   18.92 -- 19.39   &   15.72 -- 15.79   &   8.33$\times10^{-8}$   &   0.00 &   f \\
\hline
\multicolumn{10}{p{0.95\textwidth}}{$^1$ $\rho$ is the number of objects per target M dwarf} \\
\multicolumn{10}{p{0.95\textwidth}}{$^2$Total number of objects $0.05<E<0.15$}\\
\end{tabular}
\end{center}

    \end{center}
    \end{table*}

    \subsection{Chance alignment of M giants}
    \label{ch4_section_chance_align_mgiants}

    For M giants we downloaded all Milky Way stars from the {\it 10th version of the Gaia Universe Model Snapshot} \citep[GUMS-10, Milky Way stars in {\sc gums.mw}][]{Robin2012}\footnote{Accessed online at \url{http://dc.zah.uni-heidelberg.de/\_\_system\_\_/dc\_tables/show/tableinfo/gums.mw}}. The {\sc gums.mw} catalogue gives simulated stellar properties such as spectral type, luminosity class and distance, as well as the predicted {\it Gaia} $G$ magnitude to simulate objects present in the future {\it Gaia} data releases.

    From this catalogue we selected all the M giants (selecting M spectral type stars and luminosity classes {\sc i}, {\sc ii} and {\sc iii}) with galactic latitude, $b> 40\degs$, this left 4 966 M giants . We selected those with $G < 9 $ (to select a complete sample at known distance) that left 3022 M giants in our M giant sample.

    To work out a density we needed a maximum distance a $G < 9$ M giant can be observed at (and thus needed the absolute magnitude of M giants in $G$ band). Converting $G$ in to $M_{G}$ (via $M_{G} = G - 5\text{log}_{10}(\text{distance}) + 5$) and taking the faintest possible $M_{G}$ value for our M giant sample (thus the worst case scenario for our density) we estimated the maximum distance {\it Gaia} could detect M giants to was 6983 pc. Thus the density, of M giants per parsec in a spherical cone of radius, $R = 6983 pc$, $\rho_{\text{M giant}}$ is 1.1866$\times^{-8} \text{\text{M giant}} \text{ pc}^{-3}$ (using \refequ{ch4_equation_vol_sphericalcone}).

    To calculate the number of chance alignments of M giants with one of our M dwarfs we needed the maximum distance we could detect M giants out to in 2MASS/WISE. For this we needed the absolute magnitude of M giants . Using \refequ{ch4_equation_bj} \citep{Smart2016}\footnote{Equations from \citet{Smart2016} and use transformations from \url{http://www.astro.ku.dk/~erik/Tycho-2/} and \citet{Jordi2006}} we calculated an equation for $J$ in terms of G and \JK  where we take the \JK values for M giants from \citet[table 3][]{Straizys2009}.

	\begin{equation}
	\label{ch4_equation_bj}
	\begin{split}
	B_J = & \, J + 4.9816 - 0.38945670\JK \\
	R_F = & \, J + 2.6997 - 0.46257863\JK \\
	\, M = & \,R_F - J = 2.6997 - 0.46257863\JK \\
	\, G = & \,R_F + 0.0045 + 0.3623(B_J-R_F) - 0.1783(B_J-R_F)^2 \\
	& + 0.0080(B_J-R_F)^3 \\
	\, L = & \,B_J - R_F \\
	\, L = & \,J + 4.9816 - 3.8946\times10^{-1}\JK - J \\ 
	& -2.6997 + 4.6258\times10^{-1}\JK \\
	\,L = & \,7.3122\times10^{-2}\JK + 2.2819 \\
	\,J = & \, G - M - 0.0045 + 0.3623L - 0.1783L^2 + 0.0080L^3 \\
	\,J = & \,G - 3.1278\times10^{-6}\JK^3 + 6.6052\times10^{-4}\JK^2 \\ 
	& + 4.8645\times10^{-1}\JK - 2.6976 \\
	\end{split}
	\end{equation}

    For a \JK of 1.11 (average of the \JK values for M giants from Table 3 of \citet{Straizys2009}) and an absolute $G$ magnitude of -0.61 we calculate an absolute $J$ band magnitude of --2.7668. Feeding this in to the equation for distance ($d_J = 10^{-0.4(M_J - m_j)}$) where $m_j$ is the 1.25$\sigma$ limit of 2MASS$^{\ref{footnote:convert_5_to_1.25}}$ gives a distance of $\sim$200 \Mpc. This distance is far beyond the reach of the Milky Way, thus we chose to use the maximum distance observed by the excess sample, using \refequ{ch4_equation_distance_from_scale_height} taking $b_{\text{min}} = 40\degs$ and $h_z=$1200 pc as stated in the {\sc gum.mw} simulation \citep[Table 2 from][]{Robin2012}, thus giving an observed value for the excess sample of 1 867 pc.

    \begin{equation}
    \label{ch4_equation_distance_from_scale_height}
    d_{\text{max}} = \frac{h_z}{\text{sin}(b_{\text{min}})}
    \end{equation} 

    \noindent The number of chance aligned giants then comes directly from \refequ{ch4_equation_vol_sphericalcone}, where $\rho_{\text{M giant}}$ is 1.2$\times^{-8} \text{M Giants}  \text{ pc}^{-3}$, $\theta = $ 6 \arcseconds and $R = d_{\text{max}} = $1867 pc. Hence, the number of chance alignments of giant stars that are sufficiently red per M dwarf is 6.8$\times$10$^{-8}$. With 36 898 M dwarfs in our excess sample we estimate $\sim$0.003 M giant chance alignments in our excess sample.

    \subsection{Chance alignment of red galaxies}
    \label{ch4_section_chance_align_galaxies}

    For galaxies, we started with the simulation from \citet{Henriques2012}\footnote{Accessed online at \url{http://gavo.mpa-garching.mpg.de/Millennium/Help/databases/henriques2012a/database}.}. \citet{Henriques2012} use the semi-analytic models of \citet{Guo2011} that simulate the evolution of haloes and sub-haloes within them. These models are implemented on two large dark matter simulations, the Millennium Simulation \citep{Springel2005} and Millennium-Iwe Simulation \citep{BoylanKolchin2009}. 

    This gave us access to distance, $J_{AB}$, the Spizter [$4.5 \mu m$] band (also in the AB system, and assumed for simplicity to have a similar band-pass to $W2$). We chose to only count galaxies initially brighter than the WISE $W2$ 1.25$\sigma$ limit (see table 1 from \citetalias{Cook2016} where we add 1.5 to convert from 5$\sigma$ to 1.25$\sigma$) of 17 and a \JWb colour redder than 1.17 (the bluest colour our M+UCD candidates appear to be). 

    \begin{equation}
    \label{ch4_equation_cuts_to_henriques2012}
    \begin{split}
    & J_{AB} = J_{\text{Vega}} + 0.91 \\ 
    & W2_{AB} = W2_{\text{Vega}} + 3.339 \\ 
    & (J - W2)_{AB} = (J - W2)_{\text{Vega}} - 2.429 \\ 
    & [4.5 \mu m] < 20.4  \qquad \qquad J - [4.5 \mu m] > -1.259
    \end{split}
    \end{equation}

    These needed to be converted into the AB system, for $W2$ this was done using \refequ{ch4_equation_cuts_to_henriques2012} \citep[from ][]{Jarrett2011}\footnote{Accessed via \url{http://wise2.ipac.caltech.edu/docs/release/prelim/expsup/sec4\_3g.html\#WISEZMA}} and for $J$ this was done using \refequ{ch4_equation_cuts_to_henriques2012} \citep[from ][]{Blanton2007}\footnote{Accessed via \url{http://www.astronomy.ohio-state.edu/~martini/usefuldata.html}}. This led to a \JWb conversion shown in \refequ{ch4_equation_cuts_to_henriques2012}, and the cuts were then applied to the simulations by \citet{Henriques2012}. This left 11 903 galaxies in our sample of red galaxies.

    Galaxies can be red for a number of reasons, (\ie galaxies can be red because they are dusty, which reddens starlight and also emits in the infrared and via reddening due to redshift), to keep the estimation of contamination as simple as possible we use our sample of red galaxies to model the spread in \JWb observed. We took the minimum, mean and maximum values of the \JWb galaxy distribution and calculated the limiting magnitude in $J$ band (using \refequ{ch4_equation_limiting_mag}) which would give an excess of 5 per cent. This value was used as a new cut to the galaxy sample if the galaxy was brighter than the 1.25$\sigma$ limiting magnitude in 2MASS$^{\ref{footnote:convert_5_to_1.25}}$ otherwise the 1.25$\sigma$ 2MASS/WISE band limit ($18.05$/$17.00$) was used. The number of objects left gave the density of objects out to the limiting magnitude in which the galaxies would redden our M dwarfs by 5 per cent (see Table \ref{ch4_table_galaxy_results}). 

    \begin{table*}
    \caption{Table showing the limiting magnitudes to give a five per cent excess in \JWb and thus the number of chance alignments of galaxies per target M dwarf and in total for our excess sample. Note if the limiting magnitude was greater than the 1.25$\sigma$ detection limit of 2MASS or WISE, the 2MASS/WISE limit ($18.05$/$17.00$) was used instead$^{\ref{footnote:convert_5_to_1.25}}$. \label{ch4_table_galaxy_results}}
    \begin{center}
	\begin{center}
	\begin{tabular}{lcccc}
	\hline
	 & Unit & Minimum & Mean & Maximum \\
	\hline 
	\JWb & mag & 1.43 & 2.65 & 3.93 \\
	$M_{J(\text{limit})}$ & mag & 15.15 & 18.22 & 19.76 \\
	$M_{W2(\text{limit})}$ & mag & 13.72 & 15.57 & 15.83 \\
	1.25$\sigma$ 2MASS limit & mag & 18.05 & 18.05 & 18.05 \\
	1.25$\sigma$ WISE limit & mag & 17 & 17 & 17 \\
	Number of red galaxies in survey & & 39 & 898 & 1207 \\
	Density of red galaxies & Mpc$^{-3}$ & 0.001 & 0.015 & 0.020 \\
	Number of objects per target M dwarf & & 1.74$\times10^{-4}$ & 4.00$\times10^{-3}$ & 5.37$\times10^{-3}$ \\
	Total objects in with $E>0.05$ & & 10 & 143 & 175 \\
	Number of extended ($\theta > 3$ \arcseconds) & & 10 & 21 & 21 \\
	Total objects with $E>0.15$ & & 0 & 28 & 42 \\
	Total non-extended objects $0.05<E<0.15$ & & 0 & 94 & 112 \\
	\hline
	\end{tabular}
	\end{center}
    \end{center}
    \end{table*}

    There were two ways to convert this number into a number of objects chance aligned with one of our M dwarfs. Taking all galaxies to be at infinite distance one can simply divide the area of the WISE PSF (6 \arcseconds) by the area of the survey (1.4$\degs \times $1.4$\degs$); however, for consistency we also work out a density of galaxies and use the spherical cone analysis. Using the sample of galaxies we calculated the minimum, mean and maximum absolute $J$ band magnitudes and thus the minimum, mean and maximum values for the distance of these galaxies. Then using \refequ{ch4_equation_vol_sphericalcone}, we estimated a density, and the number of galaxies per M dwarf, and in the total excess sample (of 36 898 M dwarfs). All results for the minimum, mean and maximum values can be seen in Table \ref{ch4_table_galaxy_results}. 

    In our selection process \citepalias{Cook2016} any galaxy that looked extended in 2MASS or WISE was rejected as a contaminant and thus rejected from our excess sample. We thus also need to remove any galaxies in our sample that appear extended in 2MASS or WISE and thus already rejected from amongst our M dwarfs. To do this we use the hydrodynamic cosmological simulation from Fig. 3 of \citet{Naab2009} to define a relationship between redshift, $z$ and extent of the galaxy (see \refequ{ch4_equation_extent_to_logz}).

    \begin{equation}
    \label{ch4_equation_extent_to_logz}
    \left[\frac{\text{extent}}{\text{kpc}}\right] = -1.1551\text{ log}_{10}(z) + 1.2985 \qquad \qquad \text{correlation} = -0.97
    \end{equation}

    However one must be careful in converting extent of a galaxy as the angular size varies as a function of redshift (see \refequ{ch4_equation_deceleration_parameter} from equations 7.33, 7.37, 7.31 and 7.11 \citealt{Ryden2003}),

    \begin{equation}
    \label{ch4_equation_deceleration_parameter}
    \begin{split}
    & \theta = \left[\frac{\text{extent}}{d_A}\right]^{\text{rad}} \\ 
    & d_A = \frac{d_L}{(1+z)^2} \\ 
    & d_L \approx \frac{c}{H_0}z\left( 1 + \frac{1+q_0}{2}z \right) \\ 
    & q_0 = \Omega_{r,0} + 0.5 \Omega_{m,0} - \Omega_{\Lambda,0}
    \end{split}
    \end{equation}

    \noindent where $d_A$ is the angular diameter distance, $d_L$ is the luminosity distance, $z$ is the redshift, $\theta$ is in radians, $H_0 = $72 km s$^{-1}$ Mpc$^{-1}$, $q_0 = $ -0.55 is the deceleration parameter for $\Omega_{\Lambda,0} =$ 0.7, $\Omega_{m,0} =$ 0.3 and $\Omega_{r,0} = 0$ for a nearly flat universe. Combining the equations in \refequ{ch4_equation_deceleration_parameter} gives an equation for $\theta = \theta(z)$. Taking the PSF of 2MASS as 3 \arcseconds, this is equivalent to galaxies of redshift smaller than 0.05 as being possibly extended and thus already rejected.

    Our M+UCD candidates occupy an excess region between 0.05 and 0.15 in colour excess of \JWb. Therefore, we also subtract off those objects that have an excess greater than 0.15 to give a final number of objects. Hence the number of chance alignments is ~4$\times10^{-3}$ per M dwarf. With 36 898 M dwarfs in our excess sample we estimate a worst case scenario of between 94 and 112 of our excess sample may be reddened by chance alignments with red galaxies.

    \subsection{Chance alignment from random offsets}
    \label{ch4_section_chance_alignment_from_random_offsets}

    Another way we gauge possible contamination from red objects was to randomly offset our excess sample by 2\degs at random angles. This movement to a random location should simulate the possibility of finding a chance aligned object. We then cross-matched these offset points with WISE (out to 6 \arcseconds totalling 3 073 of 36 898 matches) and with 2MASS (out to 3 \arcseconds totalling 464 of the 3 073 matches). We were then able to work out \JWb colour of these objects. From this we added the object back to our M dwarfs \JWb and thus were able to calculate the objects colour excess. Of the 464 objects that had a random object with both a WISE and 2MASS detection, 105 had a positive non-zero excess and 38 had an excess between 0.05 and 0.15 (equivalent to our improvement contour constraints). Thus a total of 0.285 per cent and 0.103 per cent of our M dwarf in our excess sample had chance alignments (out of the total 36 898). This means out of our 1 082 M+UCD candidates we would have 105 objects (9.70 per cent) due to chance alignments that would produce a positive non-zero excess and 38 objects (3.51 per cent) that would produce an excess than matched our contour criteria and be selected by our approach. Thus we can expect a contamination from chance alignments of around 3.5 per cent and no worse than $\sim$9.5 per cent.

    \subsection{Visual inspection of the M+UCD candidates}
    \label{ch4_section_visual_inspection}

    As part of our reduction in contamination we visually inspected\footnote{This was achieved by using the {\sc Python} module {\sc Astroquery SkyView} \url{http://dx.doi.org/10.6084/m9.figshare.805208}} our M+UCD candidates in SDSS ($g$, $r$), in 2MASS ($J$, $H$), in WISE ($W1$, $W2$ and $W3$) and where possible in UKIDSS $J$ \citep{Lawrence2007, Lawrence2013} and the DSS2 red band. We flagged any object that were obviously blended by a red galaxy, by a diffraction spike from a bright nearby star, both of that are obvious contamination. We also flagged any object blended by a nearby object, these are also probably contamination but are not removed from our catalogue completely (due to the unknown contribution such an object gives to our M dwarf). Of our 1 082 M+UCD candidates, we found:

    \begin{itemize}
    \item 161 (14.88 per cent) as having nearby sources within the size of the WISE PSF and being possible blends;
    \item 29 (2.68 per cent) as having identifiable nearby galaxies as possible blends;
    \item 3 (0.28 per cent) as having possible contamination from the diffraction spikes of nearby bright stars;
    \item 11 (1.02 per cent) as having other problems (\ie no images or too faint to identify in one or more of the images).
    \end{itemize}

    These number seem high; however although any of these blended sources may contribute to the reddening of our M dwarfs they also might not (since the effects of the blend on the data quality are not known directly) and thus we use this number as a rough estimate of possible blended contamination. The visual inspection of the UKIDSS images (143 of the 1 082 had UKIDSS images) also showed 27 close blends 25 of these nearby objects have UKIDSS photometry (obtained by cross-matching with {\itshape UKIDSS Large Area Survey}, UKIDSS LAS, \citealt{Lawrence2007, Lawrence2013}). We then located any UKIDSS source which was within the WISE PSF and found 31 other sources around our 25 objects with UKIDSS photometry. 19 of these were flagged as galaxies using the {\sc PGalaxy}\footnote{{\sc PGalaxy} is calculated by combining individual detection classifications in the source merging process see \url{http://wsa.roe.ac.uk/www/gloss\_p.html\#lassource\_pgalaxy } for the definition} flag greater than 0.5, however as some of these nearby sources are blended with our M dwarfs they may be misclassified as galaxies.

    To gauge an upper limit on how many galaxies might be in our M+UCD candidates we cross-matched our M+UCD candidates with UKIDSS LAS (again within the WISE 6 \arcseconds PSF), 247 had matches with our 1082 M+UCD candidates. Around these 247 M dwarfs were 175 sources detected within the WISE PSF, of which 85 were flagged as galaxies ({\sc PGalaxy}$>$0.5). Thus an upper limit on the number of galaxies would be 372 (34.4 per cent), however as we discussed in Section \ref{ch4_section_chance_align_galaxies}, one needs to take into account many of these galaxies will not have the correct \JWb to give a colour excess that could mimic an unresolved UCD companion.

\end{appendix}

\bsp   
\label{lastpage}
\end{document}